\documentclass[%
 aip,
 amsmath,amssymb,
 reprint,%
]{revtex4-1}
\usepackage{xcolor}
\usepackage{graphicx}% Include figure files
\usepackage{dcolumn}% Align table columns on decimal point
\usepackage{subfigure}
\usepackage{bm}% bold math
\usepackage{braket}
\usepackage[utf8]{inputenc}
\usepackage[T1]{fontenc}
\usepackage{mathptmx}
\usepackage{etoolbox}
\usepackage{caption}
\usepackage{textcomp, gensymb}

\usepackage{float}
\usepackage{xcolor}

\makeatletter
\def\@email#1#2{%
 \endgroup
 \patchcmd{\titleblock@produce}
  {\frontmatter@RRAPformat}
  {\frontmatter@RRAPformat{\produce@RRAP{*#1\href{mailto:#2}{#2}}}\frontmatter@RRAPformat}
  {}{}
}%
\makeatother

\usepackage{tikz,xcolor,hyperref}

\definecolor{lime}{HTML}{A6CE39}
\DeclareRobustCommand{\orcidicon}{
	\begin{tikzpicture}
	\draw[lime, fill=lime] (0,0) 
	circle [radius=0.16] 
	node[white] {{\fontfamily{qag}\selectfont \tiny ID}};
	\draw[white, fill=white] (-0.0625,0.095) 
	circle [radius=0.007];
	\end{tikzpicture}
	\hspace{-2mm}
}

\foreach \x in {A, ..., Z}{\expandafter\xdef\csname orcid\x\endcsname{\noexpand\href{https://orcid.org/\csname orcidauthor\x\endcsname}
			{\noexpand\orcidicon}}
}
\begin{document}

\preprint{AIP/123-QED}

\title[]{Unshielded portable optically pumped magnetometer for the remote detection of conductive objects using eddy current measurements}
\author{L. M. Rushton\orcidE{}}
\homepage{Author to whom correspondence should be addressed:\\  lucasrushton@outlook.com}
%\homepage{The authors to whom correspondence may be addressed: lucasrushton@outlook.com, kasper.jensen@nottingham.ac.uk}
\affiliation{School of Physics and Astronomy, University of Nottingham, University Park, Nottingham, NG7 2RD, UK}
\author{T. Pyragius\orcidC{}}
\affiliation{School of Physics and Astronomy, University of Nottingham, University Park, Nottingham, NG7 2RD, UK}
\affiliation{Tokamak Energy, 173 Brook Dr, Milton, Abingdon, OX14 4SD, UK}
\author{A. Meraki\orcidD{}}
\author{L. Elson\orcidA{}}
\author{K. Jensen\orcidB{}}
\homepage{Author to whom correspondence should be addressed:\\  kasper.jensen@nottingham.ac.uk}
\affiliation{School of Physics and Astronomy, University of Nottingham, University Park, Nottingham, NG7 2RD, UK} 

\begin{abstract}
Electrically conductive objects can be detected using the principle of electromagnetic induction where a primary oscillating magnetic field induces eddy currents in the object, which in turn produce a secondary magnetic field that can be measured with a magnetometer. We have developed a portable radio-frequency optically pumped magnetometer (RF OPM) working in unshielded conditions with sub-pT/$\sqrt{\text{Hz}}$ magnetic field sensitivity when used for the detection of small oscillating magnetic fields, setting a new benchmark for the sensitivity of a portable RF OPM in unshielded conditions. 
Using this OPM, we have detected the induced magnetic field from aluminium disks with diameters as small as 1.5~cm and with the disks being $\sim 25$~cm from both the excitation coil and the magnetometer. When used for eddy current detection, our magnetometer achieves a sensitivity of a 2-6~pT/$\sqrt{\text{Hz}}$. We have also detected a moving aluminium disk using our RF OPM and analysed the magnetometer signals which depend on the position of the disk, illustrating the potential of high sensitivity RF OPMs for remote sensing applications.
\end{abstract}

\maketitle

\section{Introduction}

Optically pumped magnetometers (OPMs) \cite{budker_romalis_2007} are highly sensitive devices which can achieve sub-fT/$\sqrt{\text{Hz}}$ sensitivity \cite{kominis_kornack_allred_romalis_2003}. Compact and portable OPMs which work in shielded \cite{Osborne2018} and unshielded conditions \cite{limes_foley_kornack_caliga_mcbride_braun_lee_lucivero_romalis_2020, Fenici2020} have now been developed. This has accelerated areas of medical research such as in magnetoencephalography \cite{Xia2006,boto_meyer_shah_alem_knappe_kruger_fromhold_lim_glover_morris_et_al_2017, hill_boto_rea_holmes_leggett_coles_papastavrou_everton_hunt_sims_et_al_2020} and magnetocardiography \cite{Bison2009apl,alem_sander_mhaskar_leblanc_eswaran_steinhoff_okada_kitching_trahms_knappe_et_al_2015, jensen_skarsfeldt_staerkind_arnbak_balabas_olesen_bentzen_polzik_2018}, where small magnetic fields produced by the brain and heart are detected, respectively. OPMs can also detect oscillating magnetic fields with frequencies ranging from kHz to a few MHz \cite{savukov_seltzer_romalis_sauer_2005, Wasilewski2010prl, Chalupczak2012apl}. Such  ``RF'' optical magnetometers can be used for detecting electrically conductive objects\cite{Wickenbrock14ol,wickenbrock_leefer_blanchard_budker_2016}. 
Portable RF magnetometers in unshielded conditions have been developed with sensitivities to small oscillating magnetic fields as high as 19 pT/$\sqrt{\text{Hz}}$ \cite{deans_cohen_yao_maddox_vigilante_renzoni_2021}. 
Using the principle of electromagnetic induction, an excitation coil producing a primary oscillating magnetic field $\textbf{B}_{1}(t)$ induces eddy currents in the object, which in turn produces a secondary oscillating magnetic field $\textbf{B}_{\text{ec}}(t)$ that can be measured \cite{griffiths_stewart_gough_1999}. These eddy current measurements can be useful for imaging conductive objects with low conductivity \cite{feldkamp_quirk_2019,Jensen2019,deans_marmugi_renzoni_2020} including the human heart \cite{Marmugi2016scirep} with the potential of helping those suffering from heart diseases such as atrial fibrillation. Other applications include characterising  rechargeable batteries \cite{zhang_chatzidrosos_hu_zheng_wickenbrock_jerschow_budker_2021}, non-destructive testing \cite{bevington_gartman_chalupczak_2019, bevington_gartman_chalupczak_2021, deans_cohen_yao_maddox_vigilante_renzoni_2021}, and remotely detecting and localising conductive objects for security applications \cite{deans_marmugi_renzoni_2018, das_1990, vanverre_2021, elson_meraki_2022}.

In this work, we present a portable RF OPM working in unshielded conditions with sub-pT/$\sqrt{\text{Hz}}$ sensitivity to small oscillating magnetic fields. When detecting eddy currents, we use a differential technique \cite{Jensen2019} and in that case our magnetometer achieves a sensitivity of  2-6~pT/$\sqrt{\text{Hz}}$. We use this high-performance sensor to demonstrate a new benchmark for the long-range detection of conductive objects using a portable OPM. Here, long-range means that an object with a dimension $\sim a$ is detected at a far distance $r\gg a$ from both the excitation coil and the OPM. To be specific, we demonstrate detection with a good signal-to-noise ratio (SNR) of a 1.5~cm diameter aluminium (Al) disk at a distance of $\sim 25$~cm from both the excitation coil and the OPM, which exceeds the previous benchmark of a $\sim$10~cm size Al square plate being detected $\sim$10~cm away \cite{deans_marmugi_renzoni_2018} from both the OPM and the excitation coil. The fact that our RF OPM can detect metallic objects at a relatively large distance makes it promising for remote sensing applications. OPMs are promising alternatives to e.g. fluxgate magnetometers due to their superior magnetic field sensitivity.
We note that total-field OPMs, which are based on measuring the Larmor frequency $\omega_L\propto |B|$ (where $|B|$ is the magnitude of the total magnetic field), can be used for the detection of magnetic objects and have recently been mounted on an underwater glider \cite{page_2021} and on an airborne drone \cite{kolster_2022}. In contrast, RF OPMs can be used for the detection of both magnetic and non-magnetic conductive objects.
Extracting the size, location and motion of an object are important questions in the field of remote sensing.
As a step towards using RF OPMs for remote sensing, we here experimentally detect an aluminium disk moving along a linear path using our single RF OPM. We analyse the RF OPM response in order to extract two spatial components of the induced magnetic field which are correlated with the position of the disk along its path. 
Additional work would be needed to fully demonstrate the potential of high sensitivity RF OPMs for remote sensing, for example by simultaneously recording data from multiple RF OPMs, by placing one or more RF OPMs on a moving platform, and developing algorithms for extracting information from the recorded signals.

\section{Methods}

\subsection{Portable OPM design}

\begin{figure*}
\centering     
\subfigure[]{\label{fig:PhotoOfSetup}\includegraphics[width=10.5cm]{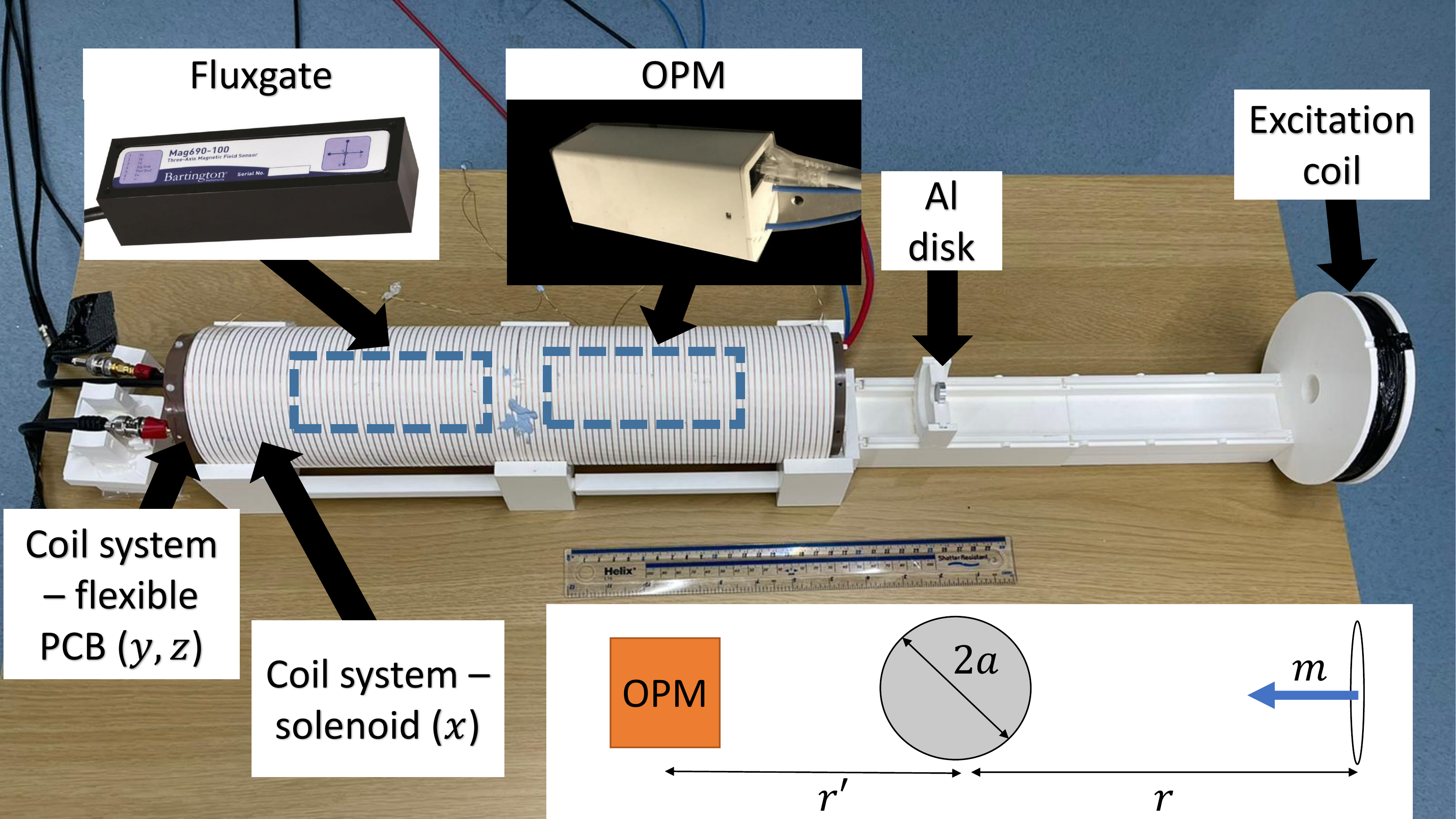}}
\subfigure[]{\label{fig:OPM_Portable_Inkscape}\includegraphics[width=6.5cm]{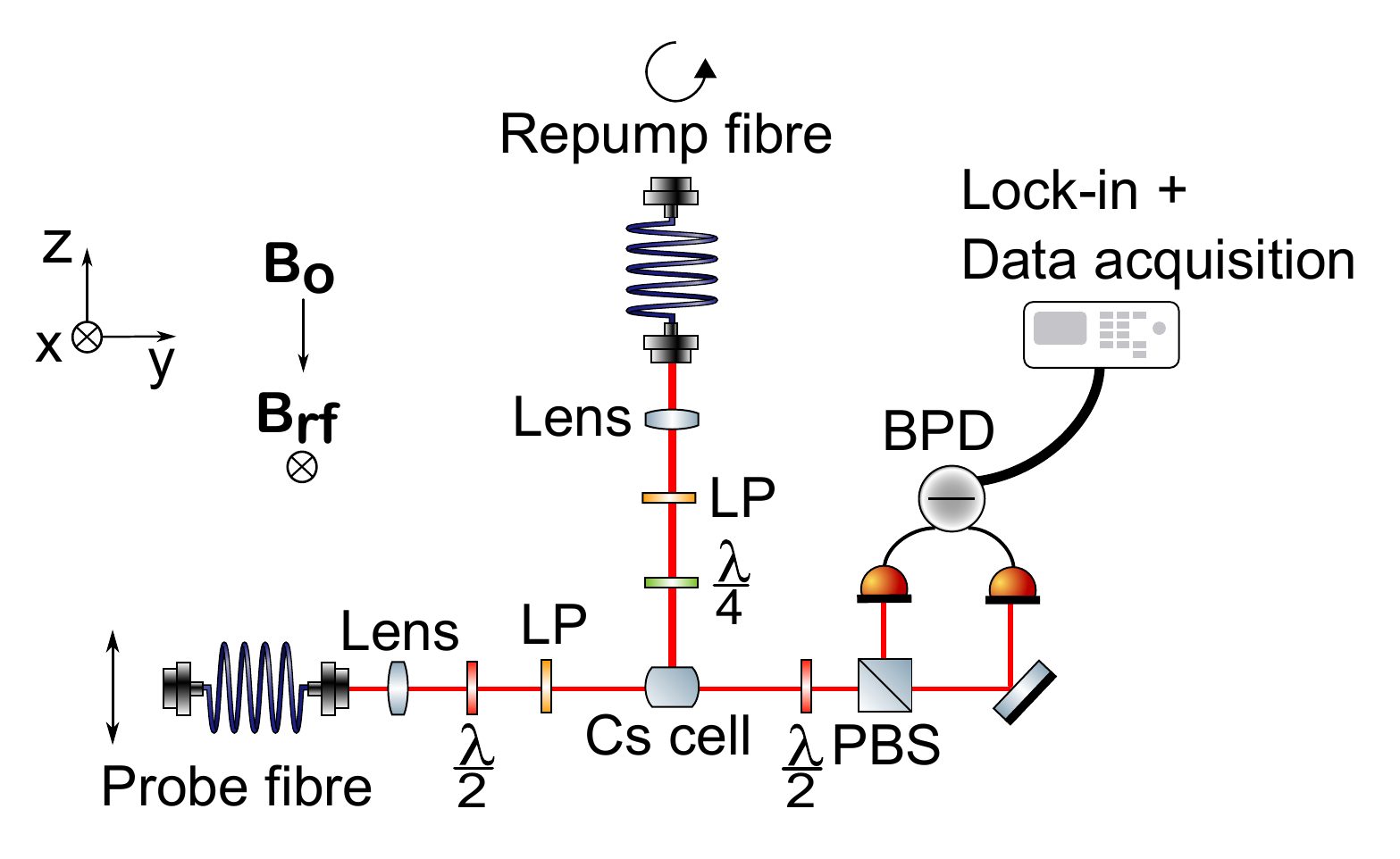}}
\caption{(a) Experimental setup. 
The OPM and fluxgate are placed in a cylindrical coil system consisting of a flexible printed circuit board (PCB) cosine-theta coils which produce homogeneous transverse magnetic fields along the $y$- and $z$-directions and also by a solenoid which produce a magnetic field along the $x$-direction. Inset: 10~cm diameter excitation coil is placed at $x=0$, the conductive object (e.g. a sphere with radius $a$) at $x=r$ and the OPM at $x=r+r'$. (b) Schematic of the portable OPM head. Components include half-wave plates ($\lambda/2$), a quarter-wave plate ($\lambda/4$), a polarising beam splitter ($\mathrm{PBS}$), linear polarisers (LP) and a balanced photodetector ($\mathrm{BPD}$).}
\end{figure*}

The experimental setup for detecting conductive objects is shown in Fig.~\ref{fig:PhotoOfSetup}. 
The setup includes an excitation coil and our unshielded portable OPM which is placed inside a cylindrical coil system. The OPM sensor head (see Fig.~\ref{fig:PhotoOfSetup} and Fig.~\ref{fig:OPM_Portable_Inkscape}) contains a cubic (5~mm)$^{3}$  caesium (Cs) vapour cell, optics, a balanced photodetector and a small compensation coil inside a 3D-printed housing. Two optical fibers provide laser light, and one cable provides electrical connections to the OPM sensor head.
The vapour cell is paraffin-coated on the inside and is kept at room temperature ($\sim 20\degree$C). The caesium atoms are optically  pumped into the $F=4$ hyperfine ground state manifold and spin-polarized in the $z$-direction by a circularly-polarised 0.1~mW pump beam resonant with the D2 $F=3~\rightarrow~F'$ transition and propagating along the $z$-direction. 
Here, $F$ and $F'$ are hyperfine quantum numbers for the caesium ground and excited states, respectively.
A static field $\textbf{B}_{0}$ is oriented along the $z$-axis. The atomic spins precess about the direction of the static field when an oscillating magnetic field $\textbf{B}_{\text{RF}}(t)$ is applied along the $x$-direction, with a maximum signal occurring when the frequency of the oscillating magnetic field $\omega_{\text{RF}}$ equals the Larmor frequency $\omega_{L}=\gamma B_{0}$, where $\gamma = 3.5$~kHz/$\mu$T is the gyromagnetic ratio for caesium. In order to detect the precession of the atomic spins, a 5~mW probe beam propagating along the $y$-axis and linearly polarised along the $z$-axis passes through the vapour cell. The probe beam is $\sim 1.8$~GHz blue-detuned from the D2 $F=4\rightarrow F'$ transition. The polarisation of the beam rotates due to the Faraday effect and hence oscillates at a frequency $\omega = \omega_{\text{RF}}$ when an oscillating magnetic field is present. The light is split into its horizontal and vertical polarisation components by a polarising beam splitter, and each beam is  incident on a balanced photodetector (BPD). The  oscillating BPD signal has an amplitude proportional to $B_{\text{RF}}$, assuming that the OPM is being operated in the low-RF amplitude regime. The signal is demodulated using a lock-in amplifier, which produces DC values for the in-phase $X$ and quadrature $Y$ signals, providing information about the amplitude $R=\sqrt{X^{2}+Y^{2}}$ and phase of the oscillating magnetic field $\textbf{B}_{\text{RF}}(t)$.

\subsection{Magnetic field stabilisation}
A stable DC field $\textbf{B}_{0}$ oriented along the $z$-axis is required for the operation of our OPM. Operating the OPM at 10.5~kHz, which is an appropriate frequency for the detection of our Al samples, requires the DC field to have an amplitude of 3.00~$\mu$T. The Earth's magnetic field is 30-60~$\mu$T and thus needs to be compensated for in order to have a stable field along the $z$-axis. A 3-axis fluxgate magnetometer (Bartington Mag690) is used to measure the ambient field and its detection point is 6.25~cm from the centre of the vapour cell. This 3-axis fluxgate has a bandwidth of 1.5 kHz and can measure magnetic fields up to $\pm$100 $\mu$T in $x$-, $y$- and $z$-directions, which makes it suitable for measuring the Earth's field as well as 50 Hz magnetic field noise. The two magnetometers are placed inside a 3D-printed cylinder, which is surrounded by flexible printed circuit board cosine-theta coils capable of producing magnetic fields along the $y$- and $z$-directions and also by a solenoid which can produce a magnetic field along the $x$-direction. The $x$-, $y$- and $z$-fluxgate outputs (100 mV/$\mu$T) are fed into the analogue inputs of a field-programmable gate array (FPGA, sbRIO-9627). 
A  proportional–integral–derivative (PID) controller implemented on the FPGA outputs a voltage to a current feedback amplifier (LT1210) and a current is sent through the coils, producing magnetic fields to cancel the Earth's field at the position of the fluxgate in the $x$- and $y$-directions, whilst keeping the $z$-static field fixed to $B_0=3.00$~$\mu$T. Without the PID in place, the 50 Hz noise measured by the fluxgate along the $z$-axis is $\sim 21.2$~nT$_{\text{p-p}}$ (corresponding to $\sim 74$~Hz when converting to Hz using the caesium gyromagnetic ratio). With the PID in place, the 50 Hz noise is reduced by at least an order of magnitude down to $\sim 2.5$~nT$_{\text{p-p}}$ (corresponding to $\sim 9$~Hz precession frequency), reducing the 50~Hz noise to below the linewidth of the magnetic resonance (40~Hz). Further noise reduction can potentially be achieved with the implementation of an active noise control system for magnetic fields \cite{Pyragius2021}.

\section{Characterisation of OPM}

\subsection{Unshielded conditions}

\begin{figure}
\includegraphics[width=\linewidth]{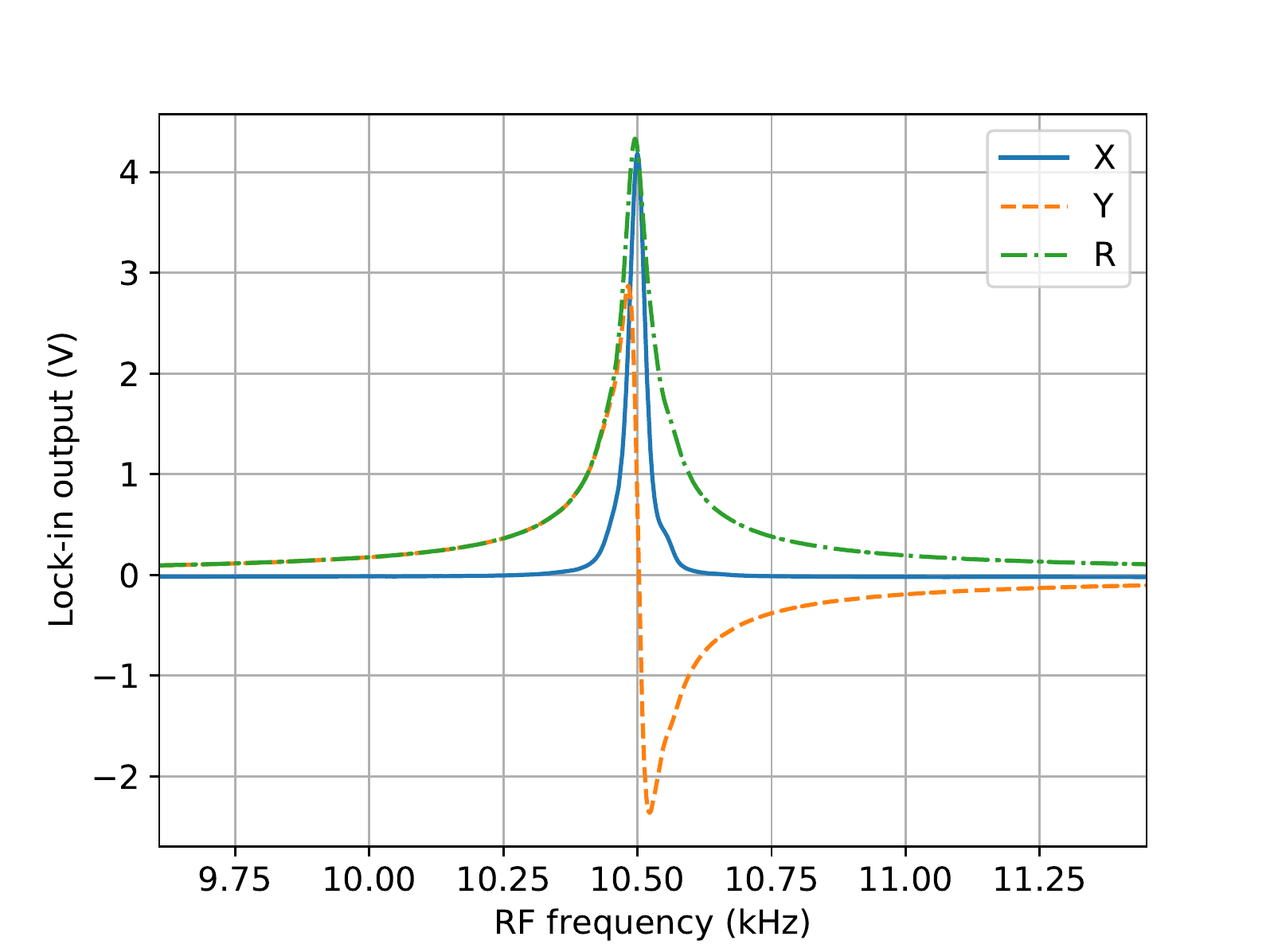}
\caption{\label{fig:FrequencySweep} Magnetic resonance. The frequency of the RF field (produced by the compensation coil) is swept between 9.75~kHz and 11.25~kHz in this data set.}
\end{figure}

\begin{figure*}
\centering     
\subfigure[]{\label{fig:B1}\includegraphics[width=5.5cm]{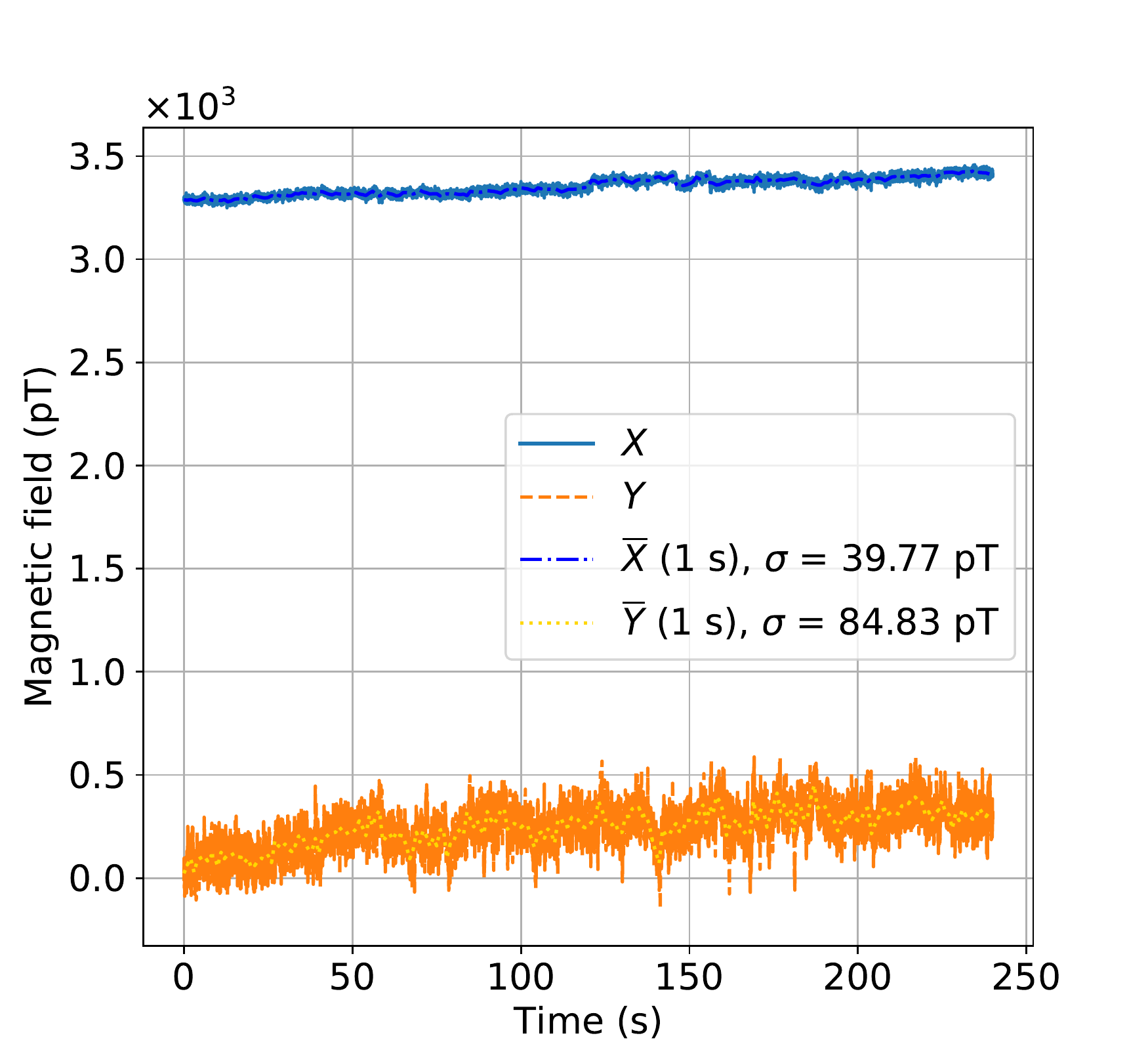}}
\subfigure[]{\label{fig:B1=B2=0}\includegraphics[width=5.5cm]{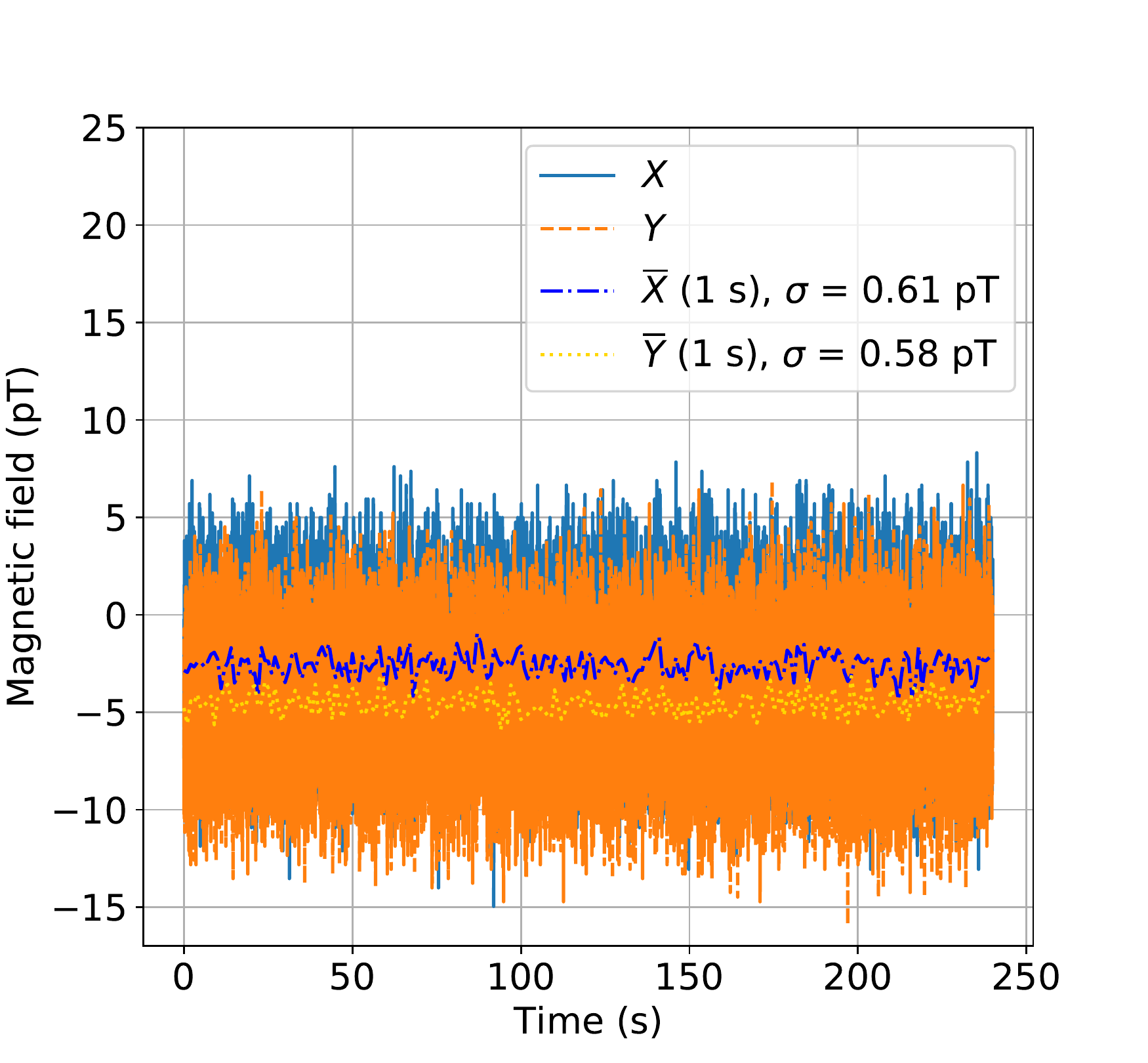}}
\subfigure[]{\label{fig:38(B1+B2)}\includegraphics[width=5.5cm]{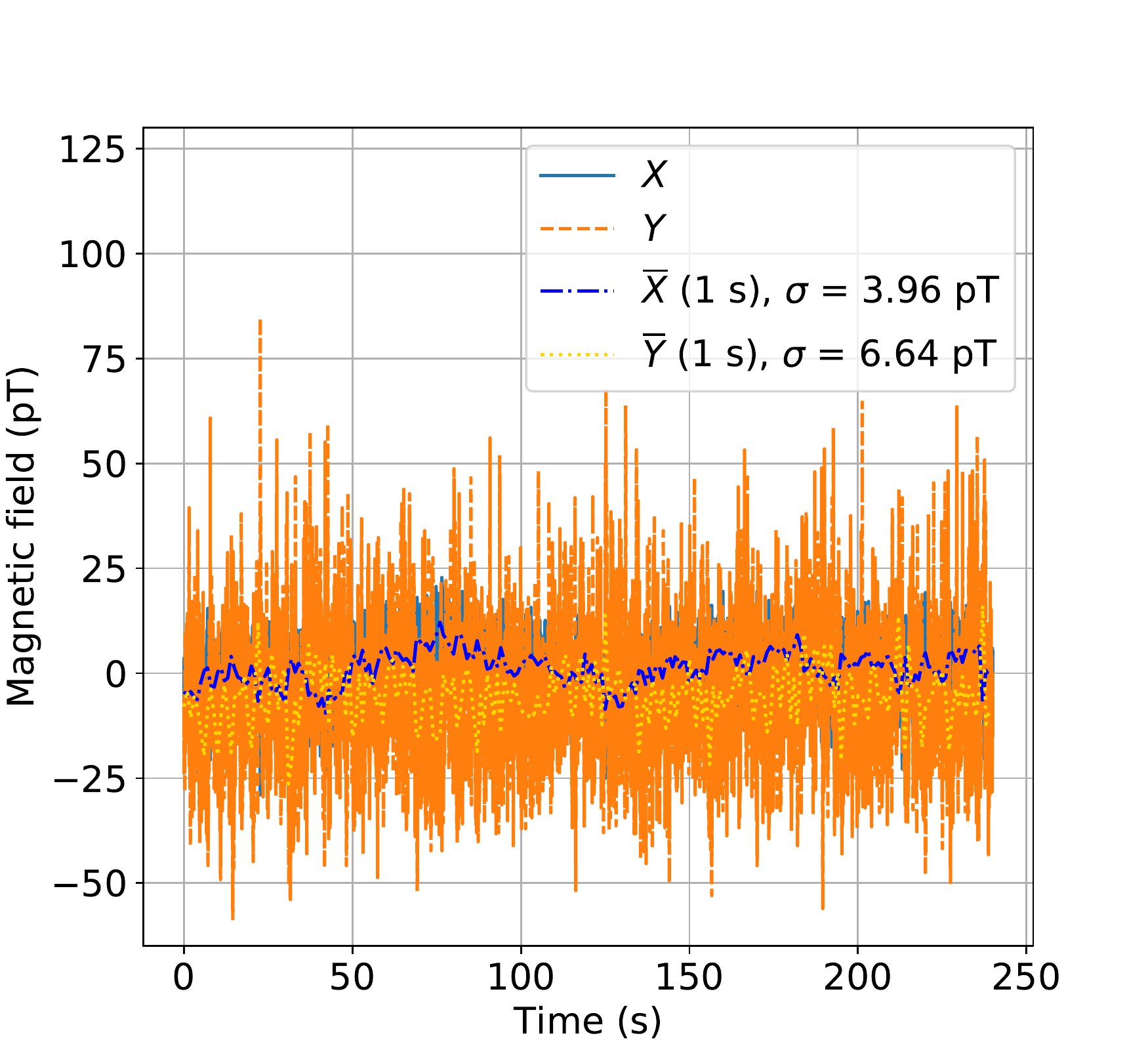}}
\caption{\label{fig:TimeTraces} Unshielded characterisation. Three sets of 240~s time traces at a frequency of 10.5~kHz. (a) Compensation coil $\textbf{B}_{2}(t)$ on with an amplitude $B_{2}=3.36$~nT$_{\text{rms}}$. (b) Excitation and compensation coils both disconnected i.e. $B_{1}=B_{2}=0$. (c) Both on with amplitudes $B_{1}=B_{2}=127.7$~nT$_{\text{rms}}$ at the position of the vapour cell such that $\textbf{B}_{1}(t,\textbf{r}_{\text{OPM}})+\textbf{B}_{2}(t,\textbf{r}_{\text{OPM}})=0$.}
\end{figure*}

The first part of the characterisation of the OPM in unshielded conditions involves measuring the magnetic resonance, which is done by sweeping the frequency of an applied oscillating magnetic field $\textbf{B}_{2}(t)$ produced by a ``compensation coil'' (5~mm diameter) inside the OPM head adjacent to the vapour cell. The amplitude of the RF field $B_{2}$ is 3.36~nT$_{\text{rms}}$
\footnote{The amplitude (in nT) of the field produced by the compensation coil was calibrated by applying DC voltages between -3~V to +3~V to the coil and measuring the shift in Larmor frequency as a function of applied voltage.}. The peak value of 4.32 V in Fig.~\ref{fig:FrequencySweep} can be used to calculate a conversion (1.285 V/nT$_{\text{rms}}$) between the lock-in amplifier output and the amplitude of the oscillating magnetic field. 
The FWHM of $X$, equivalent to the bandwidth of the OPM, is 40~Hz.

\begin{figure*}
\centering     
\subfigure[]{\label{fig:AllanDeviation_Unshielded}\includegraphics[width=8.5cm]{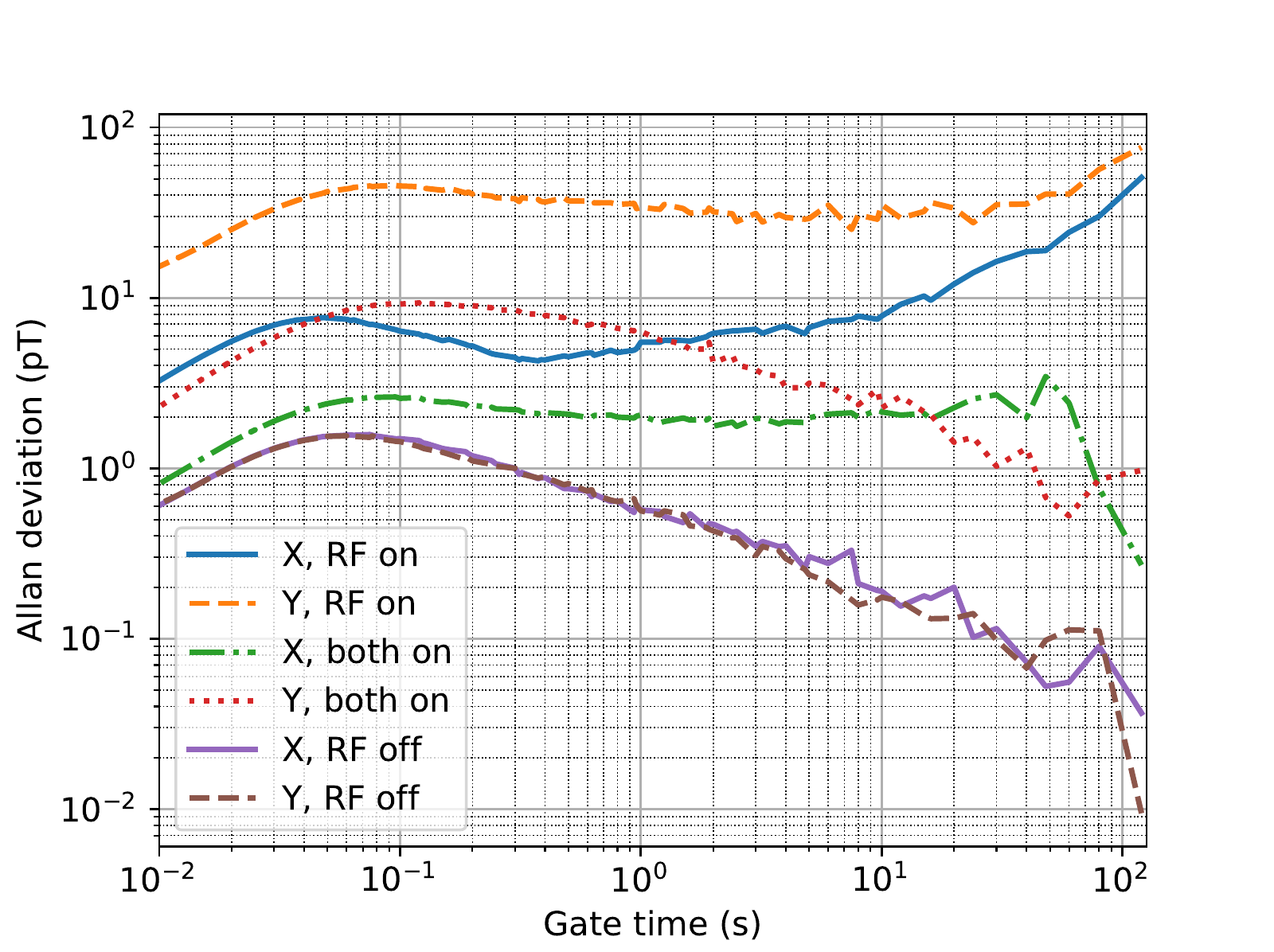}}
\subfigure[]{\label{fig:AllanDeviation_Shielded}\includegraphics[width=8.5cm]{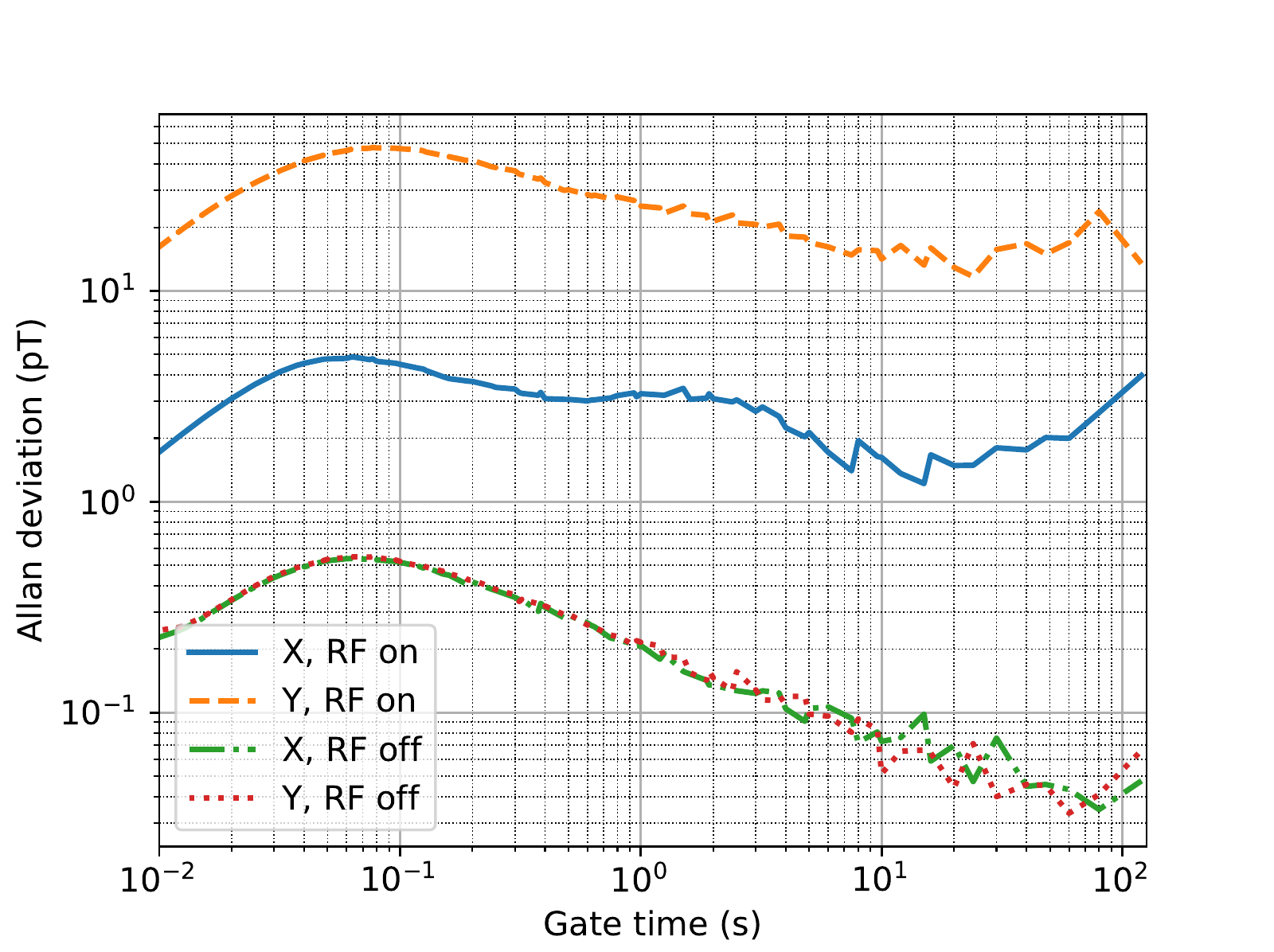}}
\caption{\label{fig:AllanDeviations} Allan deviation plots. (a) Unshielded calculations when the compensation coil is on and the excitation coil is disconnected (RF on), when both compensation and excitation coils are connected (both on) and when both coils are disconnected (RF off). (b) Shielded calculations with RF on and RF off.}
\end{figure*}

Once this step is done, the RF frequency is fixed to where there is maximum signal in $X$, which in this case is at 10.5~kHz. A 240~s time trace was taken (see Fig.~\ref{fig:B1}) for $X$ and $Y$, followed by a time trace with both coils off (see Fig.~\ref{fig:B1=B2=0}). In each time trace the averaged signals $\overline{X}$ and $\overline{Y}$ with 1~s integration times are also plotted, along with the calculated standard deviation SD of these averaged time traces. The Allan deviation of the time traces is plotted in Fig.~\ref{fig:AllanDeviation_Unshielded} (`RF on', `RF off'), which calculates the minimum detectable field $B_{\mathrm{min}}$ for different averaging (or integration/gate) times. With the RF field on, the minimum detectable field for a $\tau=1$~s integration time is $\approx 6$~pT for $X$ and $\approx 35$~pT for $Y$, while without any RF field the minimum detectable field is $\approx 0.6$~pT, i.e. there is more noise when a large RF magnetic field is applied. This could be because the applied RF field or the laser powers are not perfectly stable or because low-frequency magnetic noise gets converted to high-frequency RF noise by the OPM\cite{Jensen2019}. 
The increased noise in $Y$ compared to $X$ implies that the static field $\textbf{B}_{0}$ is noisy and a lower-noise current source or better magnetic field stabilisation should decrease the noise in $Y$. The time constant of the lock-in amplifier is 10~ms, which leads to a drop in the Allan deviation at small gate times. 
In any case, the sensitivity $\approx B_{\mathrm{min}}\sqrt{\tau}$ of an OPM is typically defined as the sensitivity to small signals, and it can therefore be stated that the sensitivity of our OPM (to small oscillating magnetic fields with frequency 10.5~kHz) is $\approx 0.6$~pT/$\sqrt{\text{Hz}}$ in unshielded conditions. The long-term stability of the OPM is also demonstrated in Fig.~\ref{fig:AllanDeviation_Unshielded}, where the minimum detectable field at an integration time of 100~s is 30-60~fT.

The characterisation of the OPM thus far has been done by applying an oscillating magnetic field $\mathbf{B}_{2}(t)$ using the small compensation coil placed inside the OPM. During eddy current measurements, the excitation coil which produces the primary oscillating magnetic field $\mathbf{B}_{1}(t)$ is also used.
We employ a differential method\cite{Jensen2019} where the amplitudes and phases of the primary and compensation fields are adjusted such that $\textbf{B}_{1}(t,\textbf{r}_{\text{OPM}})+\textbf{B}_{2}(t,\textbf{r}_{\text{OPM}})=0$ at the vapour cell position, as can be seen in Fig.~\ref{fig:38(B1+B2)} where a 240~s time trace is taken. Note that the OPM and the excitation coil are placed on opposite sides of the conductive object. This is to minimise any effects of the primary magnetic field on the OPM. When a conductive object is placed between the coils, the total oscillating field at the OPM position is then $\mathbf{B}_{\mathrm{tot}}(t, \textbf{r}_{\text{OPM}})  = \mathbf{B}_1(t, \textbf{r}_{\text{OPM}}) + \mathbf{B}_2(t, \textbf{r}_{\text{OPM}}) +  \mathbf{B}_{\text{ec}}(t, \textbf{r}_{\text{OPM}}) \approx \mathbf{B}_{\text{ec}}(t, \textbf{r}_{\text{OPM}})$, where $\mathbf{B}_{\text{ec}}(t, \textbf{r}_{\text{OPM}})$ is the secondary magnetic field generated by the conductive object. The differential technique improves the SNR and thereby allows for the detection of small objects at a distance, because it allows for the detection of the small signal  
$\mathbf{B}_{\text{ec}}(t, \textbf{r}_{\text{OPM}})$ 
on a zero background. Without the differential technique, one would measure the signal from the conductive object on top of the large primary magnetic field, i.e. 
$\mathbf{B}_{\mathrm{tot}}(t, \textbf{r}_{\text{OPM}})  = \mathbf{B}_{\text{1}}(t, \textbf{r}_{\text{OPM}}) + \mathbf{B}_{\text{ec}}(t, \textbf{r}_{\text{OPM}})$,
which for OPMs lead to non-linearities and additional noise. The measurement shown in Fig.~\ref{fig:38(B1+B2)} was done with 38 times larger oscillating fields than when just one RF coil was on (see Fig.~\ref{fig:B1}). Despite the larger applied RF fields, the Allan deviation (at a gate time of 1~s) of the OPM is around a factor of four better with both coils on than with only one coil on (2~pT/$\sqrt{\text{Hz}}$ for $X$ and 6 pT/$\sqrt{\text{Hz}}$ for $Y$ with both coils on). Taking into account the larger amplitude, this demonstrates that the differential method would give a factor of $38\times4\approx150$ improvement in SNR when detecting conductive objects. Even higher RF amplitudes would further improve the SNR.

\subsection{Shielded conditions}
The intrinsic sensitivity of the OPM was tested by placing the OPM in a magnetic shield (Twinleaf~MS-2). Time traces with the compensation coil on and off were taken, from which the Allan deviation was calculated and plotted in Fig.~\ref{fig:AllanDeviation_Shielded}. The sensitivity is 200~fT/$\sqrt{\text{Hz}}$ in shielded conditions (using $B_{\mathrm{min}}\approx 0.2$~pT for $\tau=1$~s), due to the fact that the coil system for the transverse fields did not have to be connected, thus reducing the magnetic noise at 10.5~kHz. 
Optimising the detuning of the probe beam, as well as the powers of the pump and probe beams for shielded conditions would reduce the linewidth and hence improve the sensitivity to $<$200~fT/$\sqrt{\text{Hz}}$. Heating the vapour cell to increase the density of caesium atoms would also improve the sensitivity.

\section{Eddy current measurements}

\subsection{Detection of aluminium disks with varying diameters}

\begin{figure}
\includegraphics[width=\linewidth]{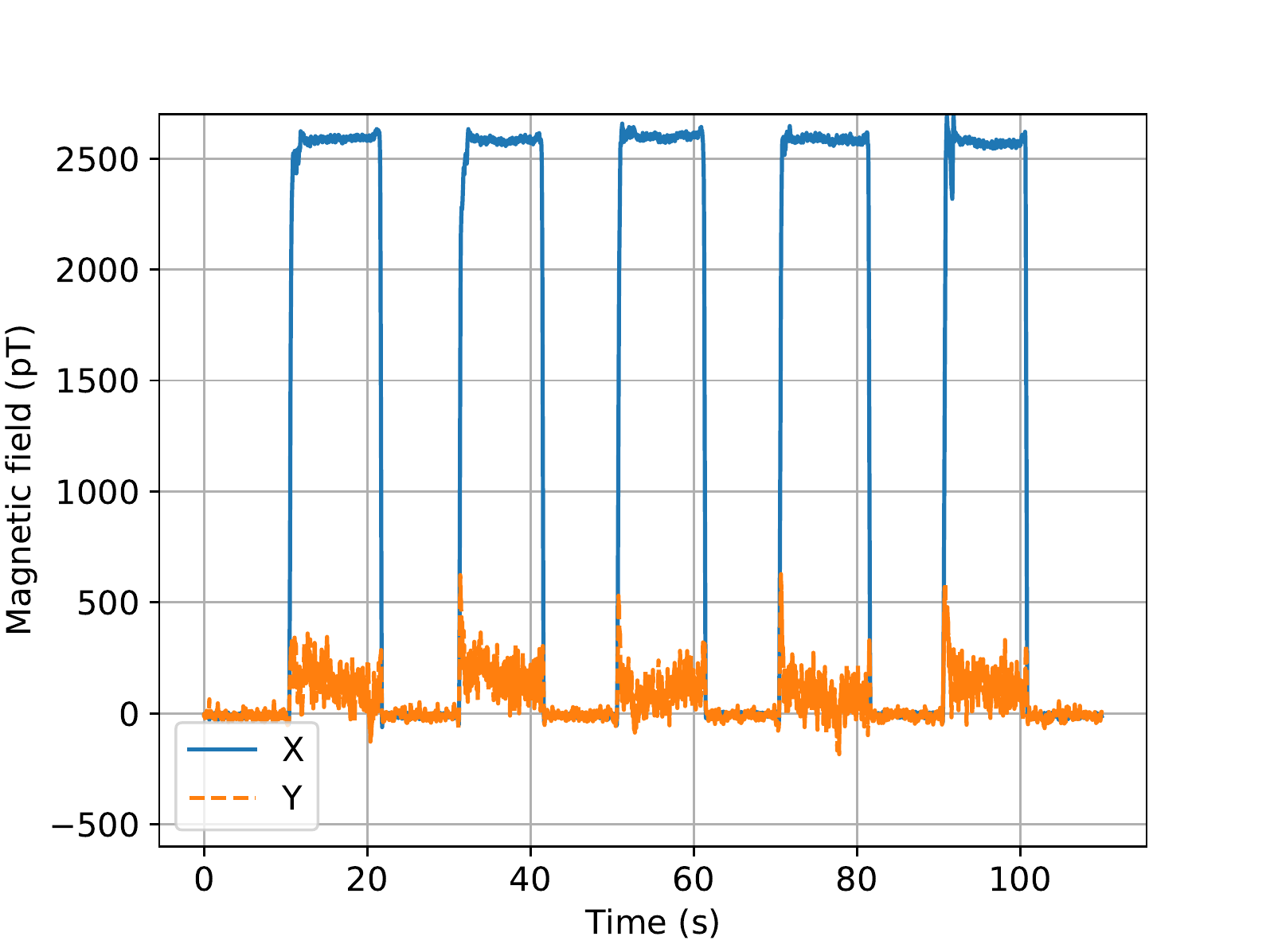}
\caption{\label{fig:ExampleEddyCurrent} Example eddy current measurement. 110~s time traces of $X$ and $Y$. 0-12~s is when the 5~cm diameter Al disk is removed, 12-22~s is when the Al disk is placed 6.4~cm from the excitation coil.}
\end{figure}

Results are now presented on the detection of Al (grade 6061 with conductivity $\sigma\sim 25$~MS/m) disks of 4~mm thickness with varying diameters (1~cm, 1.5~cm, 2~cm, 3~cm, 5~cm) in unshielded conditions using a frequency of 10.5~kHz. The excitation coil and the OPM (with the compensation coil right next to the vapour cell) are separated by 50.3~cm. One set of measurements were taken with the Al disks only 6.4~cm from the excitation coil (43.9~cm from the OPM), and a second set of measurements taken with the disks roughly halfway between the excitation coil and the OPM (23.9~cm from OPM to disk, 26.4~cm from disk to excitation coil). 
The compensation coil was used throughout these measurements.

Figure~\ref{fig:ExampleEddyCurrent} shows 110~s time traces of $X$ and $Y$ when the 5~cm diameter disk is placed 6.4~cm from the excitation coil for $\sim 10$~s (e.g. 10-22~s), then being removed for $\sim 10$~s (e.g. 23-30~s). The disk was placed in five times. The in-phase secondary magnetic field is 2600 pT and the out-of-phase secondary magnetic field is 164 pT. We observe that $|X| \gg |Y|$ meaning that the secondary magnetic field is almost completely out-of-phase (180$\degree$) with the primary magnetic field. This is expected\cite{honke_bidinosti_2018, bidinosti_chapple_hayden_2007,elson_meraki_2022} as the skin-depth in Al for a 10.5 kHz RF field is 
$\delta = 1/\sqrt{\pi f \mu_{0}\sigma}\sim 1.0$~mm which is much smaller than the 4~mm thickness of the disk. Here $f=10.5$~kHz is the excitation frequency and $\mu_0$ is the magnetic permeability of free space.

\begin{figure}
\includegraphics[width=\linewidth]{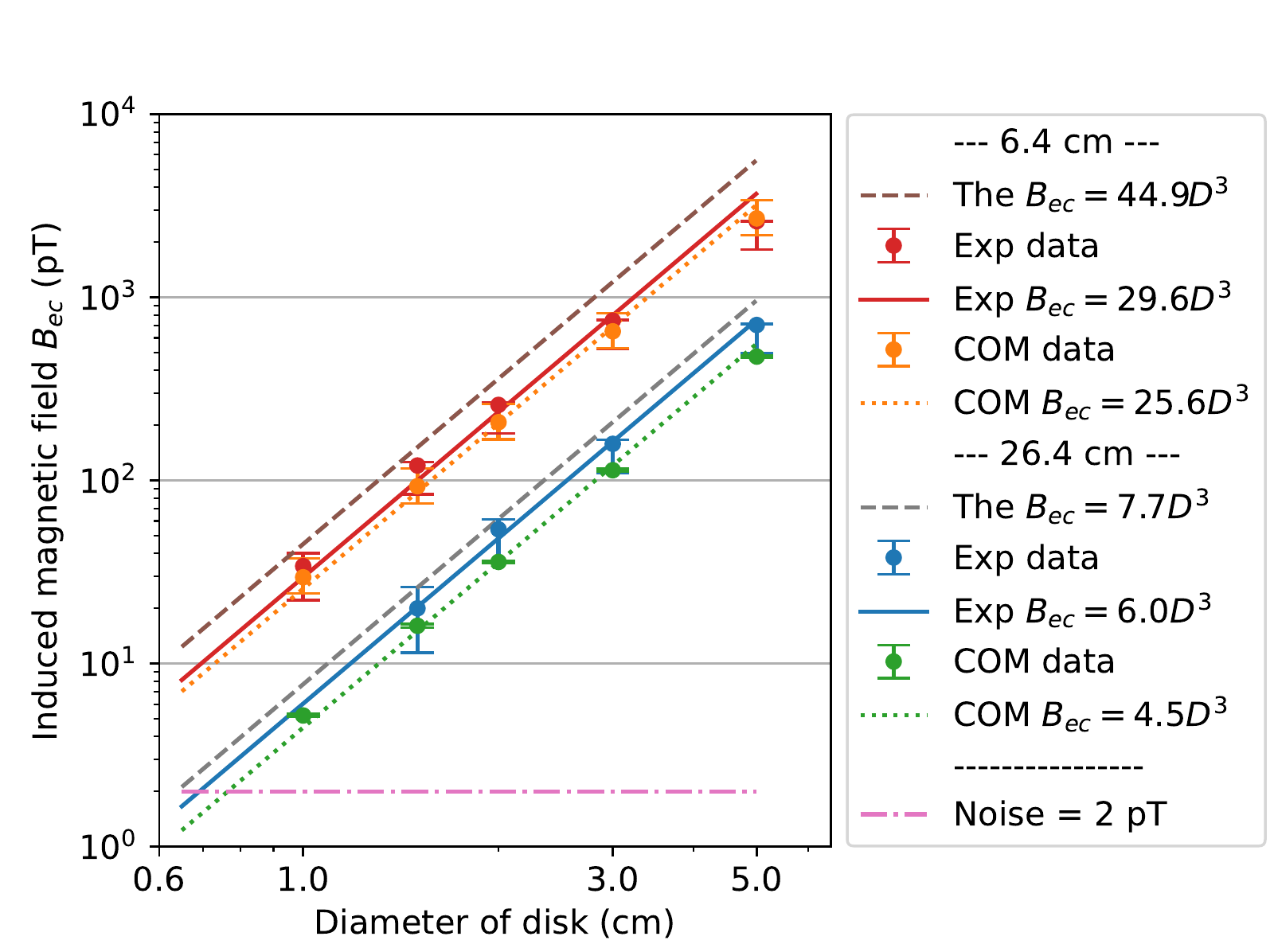}
\caption{\label{fig:InducedField_vs_Diameter} The secondary magnetic field $B_{\text{ec}}$ is plotted as a function of the Al disk diameter when (i) the Al disks are 6.4~cm from the excitation coil (43.9~cm from the OPM) and (ii) the Al disks are 26.4~cm from the excitation coil (23.9~cm from the OPM). The experimental results (``Exp'') are plotted alongside theory (``The'') curves given by Eq.~\ref{eq:bec_b1_honke2}  and results of COMSOL simulations (``COM''), together with fits to the function $B_{\text{ec}}=cD^3$.}
\end{figure}

Additional time traces when the disks are placed approximately halfway between the excitation coil and OPM are included in Appendix~\ref{sec:Appendix_EddyCurrentMeasurements}. From such time traces we can calculate the induced field in pT as a function of disk diameter for the two disk positions (see Fig.~\ref{fig:InducedField_vs_Diameter}).
We can also calculate the standard deviation SD of the 1~s integrated time traces when the object is not present, permitting for the SNR=signal/SD to be calculated for each diameter disk.
The calculated values of the SDs agree with the Allan deviation in Fig.~\ref{fig:AllanDeviation_Unshielded}, where the smallest detectable field with a 1~s integration time is $\sim 2$~pT for $X$ and $\sim 7$~pT for $Y$. When a 1.5~cm diameter disk is placed midway between the excitation coil and the OPM, the SNR is $\sim 20$ in $X$ and $\sim 2$ in $Y$, meaning that the disk is easily detectable with a good SNR.

Our experimental results are compared to analytical formulae calculated from a simple model based on the work by Honke and Bidinosti \cite{bidinosti_chapple_hayden_2007, honke_bidinosti_2018} and to the outcome of numerical simulations carried out in COMSOL. As detailed below, we find a good agreement on the scaling of the induced magnetic field with the diameter of the disks, and the predicted values for the induced field agree well with the experimentally measured ones.

In Honke and Bidinosti \cite{honke_bidinosti_2018}, $B_{\text{ec}}/B_{1}$ is calculated for all frequencies for a non-magnetic, conductive sphere with radius $a$ in a uniform magnetic field. 
In Appendix~\ref{sec:Appendix_CalculatingBec_NonMagneticSphere} we calculate the secondary magnetic field for certain positions of the excitation coil, object and OPM. If the sphere is a distance $r$ from the excitation coil and a distance $r'$ from the OPM (see inset in  Fig.~\ref{fig:PhotoOfSetup}), and the high frequency limit is considered, then 
\begin{equation}
    \frac{B_{\text{ec}}}{B_{1}} = \frac{a^{3}(r+r')^{3}}{r^{3}r'^{3}}
    \label{eq:bec_b1_honke}
\end{equation}
at the position of the OPM. If the object is exactly halfway between the excitation coil and the OPM (i.e $r=r'$), Eq.~\ref{eq:bec_b1_honke} further simplifies to 
\begin{equation}
    \frac{B_{\text{ec}}}{B_{1}} = \frac{\left(2a\right)^{3}}{r^{3}}.
     \label{eq:bec_b1_honke2}
\end{equation}
The experimental data sets in Fig.~\ref{fig:InducedField_vs_Diameter} are fitted to the function $\log \left( B_{\text{ec}} \right)=\log \left(c \right) + 3 \log \left( D\right)$, corresponding to the power law dependence $B_{\text{ec}}=cD^3$ as in Eq.~\ref{eq:bec_b1_honke2}. Here $D=2a$ is the diameter $D$ of the disks in cm. The constant $c$ is equal to $B_{1}(r+r')^{3}/(8r^{3}r'^{3})$ when $r\neq r'$ and equal to $B_{1}/r^{3}$ when $r=r'$ (see Eqs.~\ref{eq:bec_b1_honke} and  \ref{eq:bec_b1_honke2}). The fitted constant $c_{\text{exp}}$ in Fig.~\ref{fig:InducedField_vs_Diameter} when the disk is 26.4~cm from the excitation coil is 6.0~pT/cm$^{3}$, whereas the theoretical value $c_{\text{theory}}$ is 8.1~pT/cm$^{3}$ i.e. 34\% higher than $c_{\text{exp}}$ (using $r=26.4$~cm, $r'=23.9$~cm and $B_{1}=127.7$~nT$_{\text{rms}}$). When the disk is close to the excitation coil $c_{\text{exp}}=29.6$~pT/cm$^{3}$, whereas $c_{\text{theory}}=91.6$~pT/cm$^{3}$ i.e. 210\% higher than $c_{\text{exp}}$ (using $r=6.4$~cm, $r'=43.9$~cm). There is a larger discrepancy when the disk is closer to the excitation coil where the radius of the excitation coil $R_{c}$ (5~cm) is similar to the distance $r$ from the centre of the coil to the disk (6.4~cm). Eq.~\ref{eq:bec_b1_honke} and hence the calculation of $c_{\text{theory}}$ assumes that the primary magnetic field is a magnetic dipole. The primary magnetic field $B_{1}(x=r)$ at the position of the disk is a factor of $(r^{2}+R_{c}^{2})^{3/2}/r^{3} = 2.04$ smaller (see Eq.~\ref{eq:b1_close_to_excitation_coil}-\ref{eq:figureofmerit_b1closetoexcitationcoil}) if the primary magnetic field from a coil with a finite radius $R_{c}$ is used instead of the primary magnetic field from a magnetic dipole. This reduces $c_{\text{theory}}$ by a factor of 0.49 down to 44.9 pT/cm$^{3}$, around 52\% higher than $c_{\text{exp}}$. For the 26.4~cm disk position the constant $c_{\text{theory}}$ is only affected slightly as $R_{c} \ll r$, with a correction from 8.1~pT/cm$^{3}$ to 7.7~pT/cm$^{3}$, indicating a 27\% overestimation of $c_{\text{theory}}$ versus $c_{\text{exp}}$.

To investigate the discrepancy between theory and experiment further, numerical simulations of the experimental setup  were performed in COMSOL using the methods in\cite{elson_meraki_2022}. The data points from the simulations are included in Fig.~\ref{fig:InducedField_vs_Diameter}. Uncertainties in the positioning of the disks ($\pm$1~cm) were included in the error bars in the COMSOL data. The finite thickness (2~cm) of the coil and the uncertainty on the OPM position were not taken into account, although these would also contribute to uncertainties in the numerical simulations. 
With regard to experimental uncertainties,
we calculate the standard deviation of the induced field from 5 repeated measurements, see e.g. Fig.~\ref{fig:ExampleEddyCurrent}. 
Furthermore, the eddy current measurements were taken over the course of several hours ($r=6.4$~cm data followed by $r=26.4$~cm data) when the lab temperature gradually increased throughout this period of time, leading to an increased number density of caesium atoms throughout the day. Due to the room-temperature operation of this OPM, the resonance signal amplitude of the OPM increased between the beginning (4.32~V in Fig.~\ref{fig:FrequencySweep}) and end (5.56~V) of the day by $\sim$30\%. 
A temperature increase of $2.5^{\circ}$C will lead to an increase in the atomic density by 30\% \cite{Steck_2019}.
The data in Fig.~\ref{fig:FrequencySweep}-\ref{fig:TimeTraces} for the sensitivity measurements was obtained within minutes of each other at the beginning of the day and so temperature changes will have had little impact on these measurements. The calibration at the beginning of the day was used for the eddy current measurements, meaning that in fact \textit{smaller} $B_{\text{ec}}$ values were being measured than in the stated calibrated pT values in Fig.~\ref{fig:ExampleEddyCurrent} and Fig.~\ref{fig:TimeTraces_AllDiameters}. Including this uncertainty in the errorbars on the experimental data in Fig.~\ref{fig:InducedField_vs_Diameter} means that the experimental data and the COMSOL data are in agreement with each other. The differences between experiment/COMSOL and theory are most likely due to the fact that the theory is true for a solid sphere in a uniform RF field, while in the experiment/COMSOL simulations we detected a solid disk\cite{nagel_2018, nagel_correction}.

The ratio $B_{\text{ec}}/B_{1}$ measured at the OPM position can be used as a figure of merit for the remote detection of conductive objects. For the 1.5~cm diameter disk which was clearly detectable, we have
$B_{\text{ec}}/B_{1}\sim 2\cdot10^{-4}$. Using the noise level of 2~pT for $X$ from the Allan deviation calculations, the smallest detectable diameter should be around 0.7~cm (see Fig.~\ref{fig:InducedField_vs_Diameter}), leading to a ratio as small as $B_{\text{ec}}/B_{1}\sim 2\cdot10^{-5}$. For comparison, a 2~cm diameter coin (87\% Cu) is detected 7.5~cm from the excitation/sensing coil in \cite{vanverre_2021} with a good SNR, giving a ratio of $B_{\text{ec}}/B_{1}\sim2\times10^{-2}$. We are able to detect a small ratio and therefore able to detect small objects at relatively large distances for two reasons:
firstly, our OPM (at $x=r+r'$) and excitation coil (at $x=0$) are placed on opposite sides of the disk (at $x\approx r$), which means that the ratio $B_{\text{ec}}/B_{1}$ is improved by a factor of $ 1/\left[R_{c}^3/(8r^{3})\right]\sim 500$, where $R_c$ is the radius of the excitation coil (see Appendix~\ref{sec:Appendix_AdvantageDifferentialTechnique}), compared to the case where the OPM and excitation coil are co-located; secondly, by implementing the differential technique we achieved an improvement in SNR by a factor of 150.

\subsection{Detection of a moving aluminium disk}

To illustrate the potential of using RF OPMs for remote sensing, we have detected the 5~cm diameter disk as it was moved off-axis along a linear path from $y$~=~-22.5~cm to $y$~=~22.5~cm at a fixed $x$~=~6.4~cm position. The disk was moved by hand with an approximately constant velocity on an orthogonal rail (not shown) parallel to the table which was added to the setup in Fig.~\ref{fig:PhotoOfSetup} to steer the motion.
As the disk is being moved in the $x-y$ plane, $B_{\text{ec, z}}= 0$ due to symmetry. The  $B_{\text{ec, x}}$ and $B_{\text{ec, y}}$ components are in general non-zero when the disk is placed in the $x-y$ plane, however for the specific case of the object being on-axis (i.e. placed on the $x$-axis), the induced magnetic field only has a $B_{\text{ec, x}}$ component at the magnetometer position.

RF OPMs are sensitive to oscillating magnetic fields perpendicular to the direction of the static field $\textbf{B}_{0}$, which in our case are the $x$- and $y$-directions.
The measured secondary field can be written as 
\begin{equation}
\mathbf{B}_{\text{ec}}(t, \textbf{r}_{\text{OPM}}) =
\left[
B_{\text{ec,x}}(t) \mathbf{\widehat{x}} +
B_{\text{ec,y}}(t) \mathbf{\widehat{y}}
\right]
\cos\left( \omega_{\text{RF}} t + \theta \right),
\end{equation}
where $\mathbf{\widehat{x}}$ and $\mathbf{\widehat{y}}$ are unit vectors along the $x$- and $y$-directions.
For a moving disk, the amplitudes of the induced field at the magnetometer position, $B_{\text{ec,x}}(t)$ and $B_{\text{ec,y}}(t)$, will vary slowly as a function of time due to the changing position of the disk. 
Overall, the induced field is oscillating at the excitation frequency  $\omega_{\text{RF}}$ and with a phase $\theta$ which here is defined as the phase relative to the compensation field (which is 180$\degree$ out-of-phase with the primary field).
The phase $\theta$ should not depend on the position of the disk.
When the thickness $t$ of the disk is much larger than the skin depth $\delta$, or equivalently the excitation frequency $f=2\pi\omega_{\mathrm{RF}} \gg 1/\left(\pi t^2 \mu_0 \sigma \right)$, then the secondary field will be 180$\degree$ out-of-phase with the primary field \cite{elson_meraki_2022} corresponding to a phase $\theta = 0$.
From the Bloch equations describing an RF OPM \cite{Jensen2019}, one can show  that the recorded lock-in magnetometer signals are
\begin{align} \label{eqn:XYoffaxis}
X(t)  \propto   \: B_{\text{ec, x}}(t) \cos \left( \theta \right) 
- B_{\text{ec, y}}(t) \sin \left( \theta \right) & \nonumber \\
Y(t)  \propto   - B_{\text{ec, x}}(t) \sin \left( \theta \right) 
- B_{\text{ec, y}}(t) \cos \left( \theta \right) & \quad \textit{(disk off-axis).}
\end{align}
In the above we assumed that the amplitudes $B_{\text{ec,x}}(t)$ and $B_{\text{ec,y}}(t)$ vary slowly in time compared to the oscillation period $1/f$, the inverse of the magnetometer bandwidth (1/40~Hz) and the lock-in time constant of 10~ms.
From Eq.~\ref{eqn:XYoffaxis} we see that the lock-in outputs $X(t)$ and $Y(t)$ from the RF OPM depend on the $x$- and $y$-components of the induced magnetic field, $B_{\text{ec, x}}(t)$ and $B_{\text{ec, y}}(t)$ respectively, as well as its phase $\theta$.
When the object is placed on-axis, the measured secondary field only has an $x$-component and the lock-in signals are
\begin{align}  \label{eqn:XYonaxis}
X(t)  \propto   B_{\text{ec, x}}(t) \cos \left( \theta \right)  & \nonumber \\
Y(t)  \propto  - B_{\text{ec, x}}(t) \sin \left( \theta \right)  & \quad \quad \textit{(disk on-axis).}
\end{align}

\begin{figure}
\includegraphics[width=\linewidth]{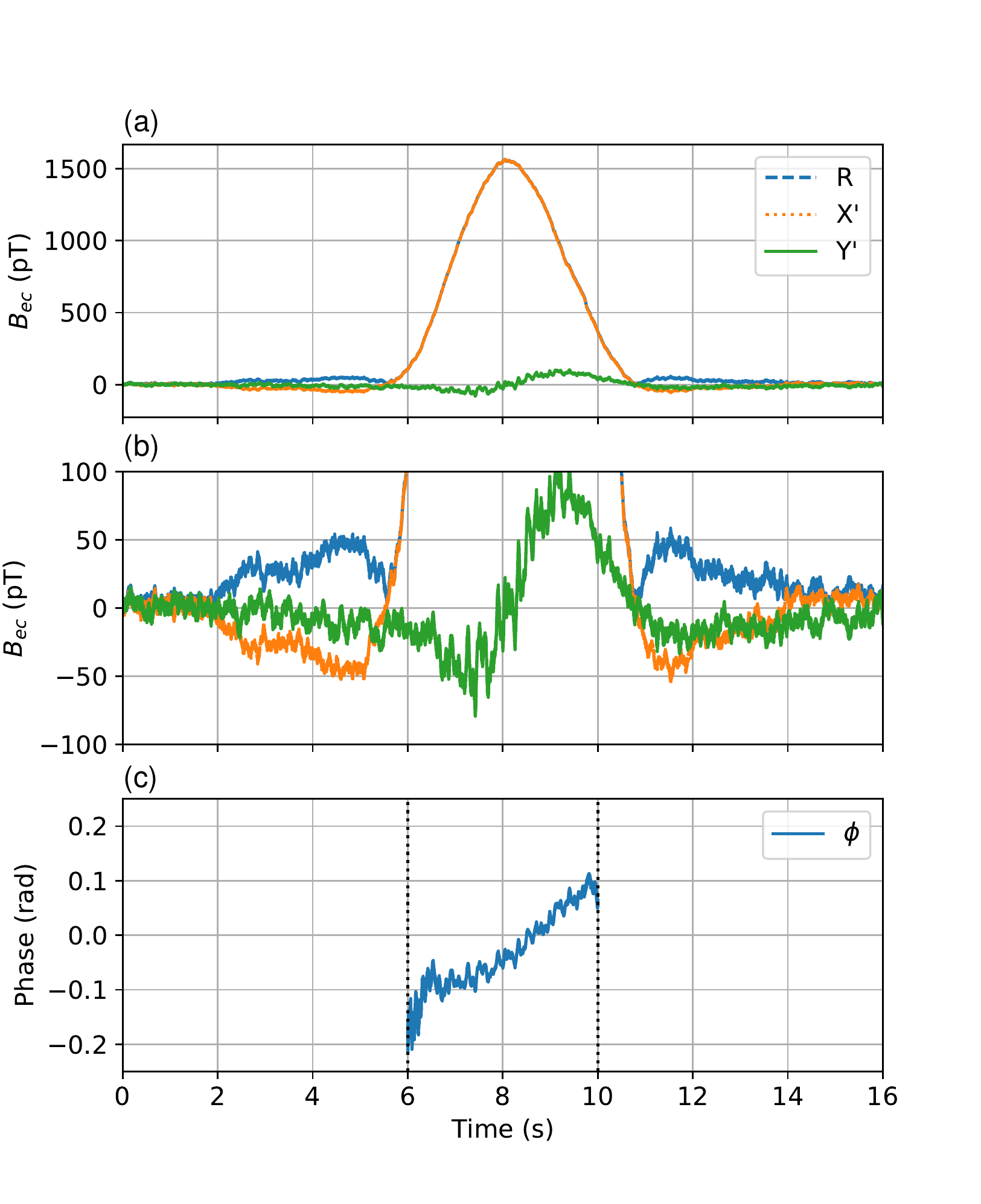}
\caption{\label{fig:OffAxis}
Off-axis example. The 5~cm Al disk is moved from $y=-22.5$~cm to $y=22.5$~cm at a distance of $x=6.4$~cm from the excitation coil. The magnitude ($R$), rotated in-phase $X'$ and quadrature $Y'$ components are plotted in (a), with a zoomed-in section in (b). The phase is plotted in (c), and the regions furthest from $y=0$ are excluded as $X$ and $Y$ become very small, making the calculated phase less insightful.
}
\end{figure}

Figure~\ref{fig:OffAxis} shows the magnitude $R=\sqrt{X^2+Y^2}$ and phase $\phi=\arctan\left(Y/X\right)$ of the recorded signals when the disk is moved off-axis along the described linear path. 
The largest signal in $R\approx1560$~pT occurs at $t=8.1(0.2)$~s  when the disk is 
located on-axis, i.e. at position $x=6.4$~cm and $y=0$. At that point, the recorded phase $\phi = -\theta \approx -0.04(0.02)$~rad  $=-2(1)\degree$ is close to zero as expected.
We note that in the experiment the RF field could be slightly detuned from the atomic resonance due to small drifts in the bias magnetic field, which would lead to a small phase offset as well.

In our experiment the phase $\theta$ is close to zero. In a more general situation, however, the phase $\theta$ will be non-zero and will depend on the object's size and shape, its electrical conductivity and magnetic permeability, and the excitation frequency \cite{elson_meraki_2022}. However, the phase $\theta$ should not depend on the position of the object. For the localisation of an object, it can therefore be useful to remove the dependence on the phase $\theta$ by rotating the lock-in outputs $X$ and $Y$ 
 from the RF OPM (see Eq.~\ref{eqn:XYoffaxis}) by the angle $-\theta$, giving rise to the rotated variables
\begin{align} 
X'(t) \propto B_{\text{ec, x}}(t) & \nonumber \\
Y'(t) \propto -B_{\text{ec, y}}(t) & .
\end{align}
Based on the geometry/symmetry of our experimental setup and the fact that the aluminium disk is moving parallel to the $y$-axis, we expect that 
$B_{\text{ec, x}} \propto X'$ is symmetric around $y=0$ (equivalent to 8.1(0.2)~s in Fig.~\ref{fig:OffAxis}) as a function of $y$-position, and that 
$B_{\text{ec, y}} \propto Y'$ is asymmetric around $y=0$ as a function of $y$-position.  Within reasonably good agreement, we find experimentally (see Fig.~\ref{fig:OffAxis}) that $X'$ is symmetric and $Y'$ is asymmetric, as expected. Any small discrepancies are expected to be due to small positioning errors/misalignment.
We also note that for every position of the disk along its particular linear path there is a corresponding unique ($X'$, $Y'$) value measured by the RF OPM, meaning that the position of the disk along its particular linear path and the direction of motion can be extracted. The velocity of the object was calculated to be 
$v \sim 0.45~\text{m} / 16~\text{s}\sim 0.028$~m/s. The limitation on the maximum detectable velocity is set by the bandwidth of the OPM (40 Hz), as this is on a slower time scale than the lock-in time constant and oscillation period. Assuming at least 20 data points would be needed to produce similar data to Fig.~\ref{fig:OffAxis}, the maximum velocity for this configuration would be on the order of 0.45~m / (20 $\times$ 1 / (40 Hz)) = 0.9~m/s. Using a commercial fluxgate magnetometer (Bartington Mag690) with a 1 kHz bandwidth would allow for velocities as high as 22.5~m/s to be detected with this configuration \cite{elson_meraki_2022}.

Our method of detecting conductive objects using RF OPMs can potentially be extended to localising unknown conductive objects moving along arbitrary paths. As a single RF OPM only provides two measurements $X(t)$ and $Y(t)$ at each instance of time, more RF OPMs would be needed to uniquely determine the position of the object in real time. Furthermore, one would need to develop algorithms for extracting the location of the object based on the recorded data. 
Also, localisation of stationary conductive objects using one or more RF OPMs could be done by placing the RF OPMs on a moving platform and recording data while the platform is moving over some area.

\section{Conclusions}
In conclusion, we have developed a portable sub-pT/$\sqrt{\text{Hz}}$ (when $B_{1}(t) = B_{2}(t) \sim 0$) radio-frequency optically pumped magnetometer (RF OPM) 
working in unshielded/ambient conditions, 
setting a new benchmark for the sensitivity of a portable RF OPM in unshielded conditions. 
Using electromagnetic induction, we have demonstrated remote detection of electrically conductive objects far from both the excitation coil and the magnetometer. We detected a $2a=1.5$~cm diameter aluminium disk at a distance of $r\sim 25$~cm from both the OPM and the excitation coil i.e. at a distance $r\gg a$ much larger than the object size. This detection distance could be further extended using larger primary magnetic fields or by improving the sensitivity of the OPM, which had a sensitivity of 2~pT/$\sqrt{\text{Hz}}$ during the eddy current measurements when $\textbf{B}_{1}(t)+\textbf{B}_{2}(t) = 0$ and $B_{1}(t)=B_{2}(t)\gg 0$. 
To illustrate the potential of high sensitivity RF OPMs for remote sensing applications, we detected a moving aluminium disk using our RF OPM.
We analysed the magnetometer signals to extract two spatial components of the induced magnetic field which depend on the position of the disk. Using this principle with multiple OPMs and an extraction algorithm should allow for the location and motion of conductive objects to be determined in the future.

\begin{acknowledgments}
This work was supported by the UK Quantum Technology Hub in Sensing and Timing, funded by the Engineering and Physical Sciences Research Council (EPSRC) (Grant No. EP/T001046/1), the QuantERA grant C’MON-QSENS! by EPSRC (Grant No. EP/T027126/1), the Nottingham Impact Accelerator/EPSRC Impact Acceleration Account (IAA),  the Novo Nordisk Foundation (Grant No. NNF20OC0064182), and Dstl via the Defence and Security Accelerator (www.gov.uk/dasa). We thank Thomas Fernholz for reading and commenting on the manuscript.
\end{acknowledgments}

\section*{Data Availability Statement}
The data that support the findings of this study are available from the corresponding author upon reasonable request.

\appendix

\section{Eddy current measurements}
\label{sec:Appendix_EddyCurrentMeasurements}
\begin{figure*}
%\centering     %%% not \center
\subfigure[]{\label{fig:5cm_diameter}\includegraphics[width=8.5cm]{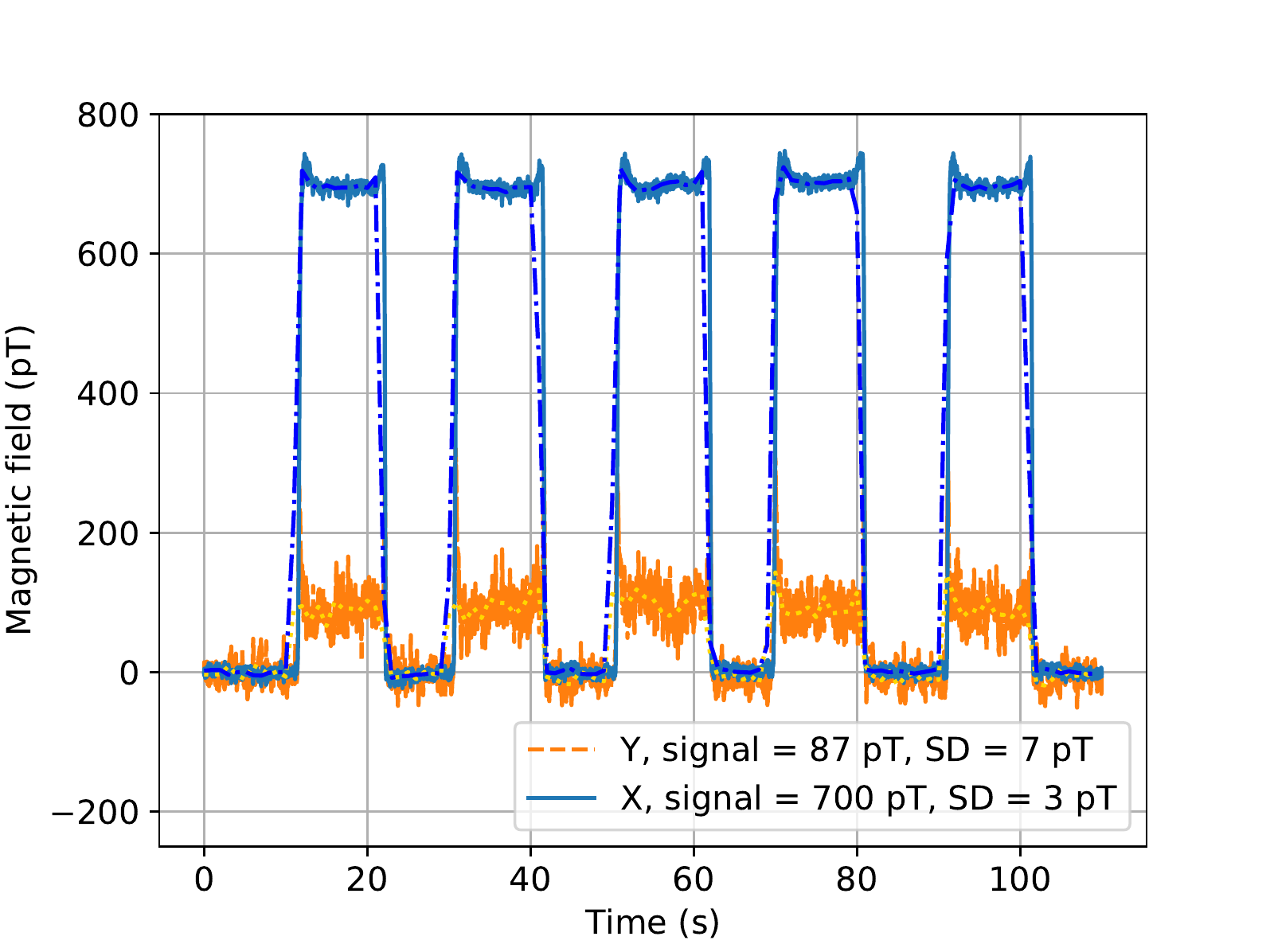}}
\subfigure[]{\label{fig:3cm_diameter}\includegraphics[width=8.5cm]{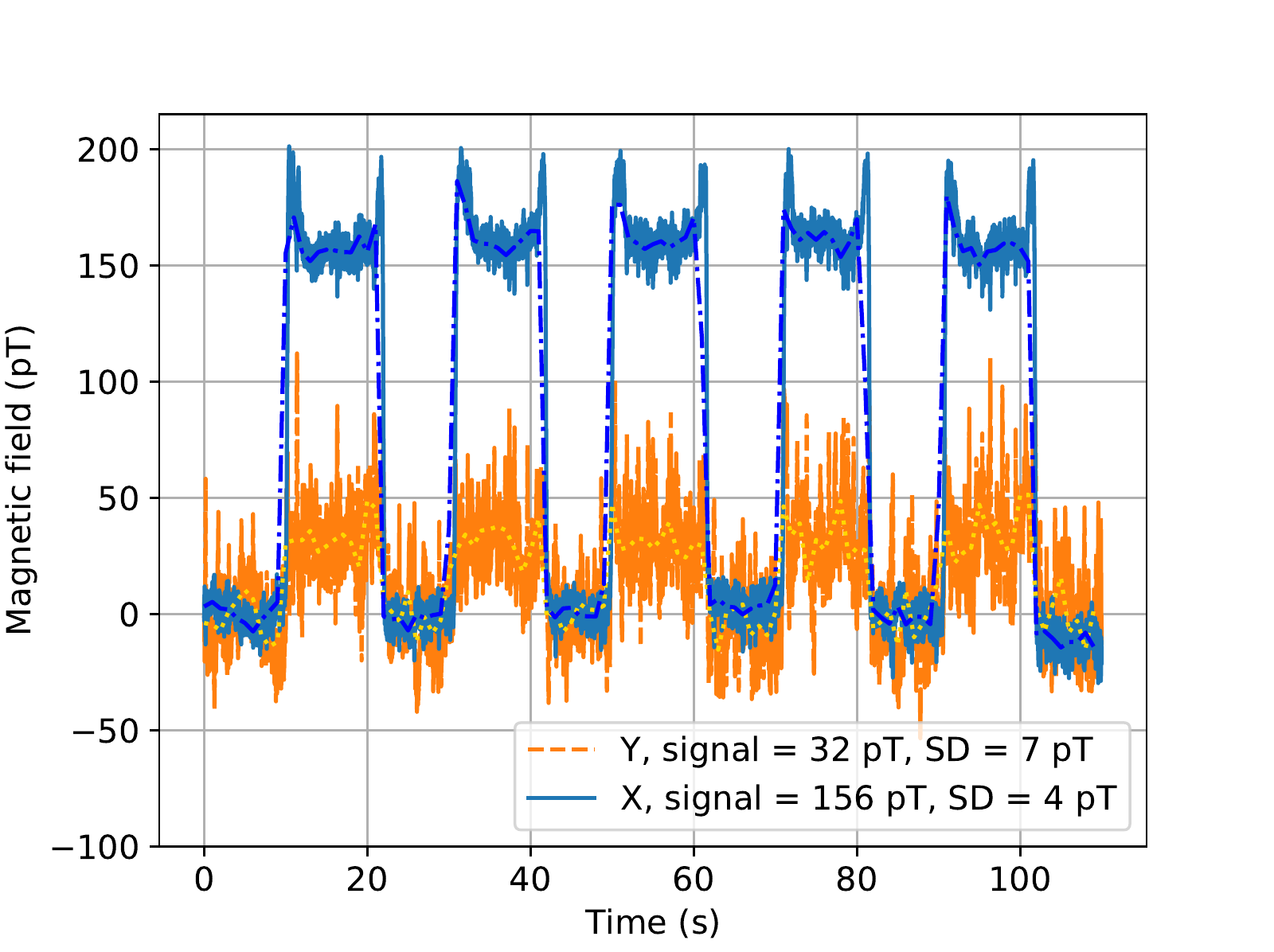}}
\subfigure[]{\label{fig:2cm_diameter}\includegraphics[width=8.5cm]{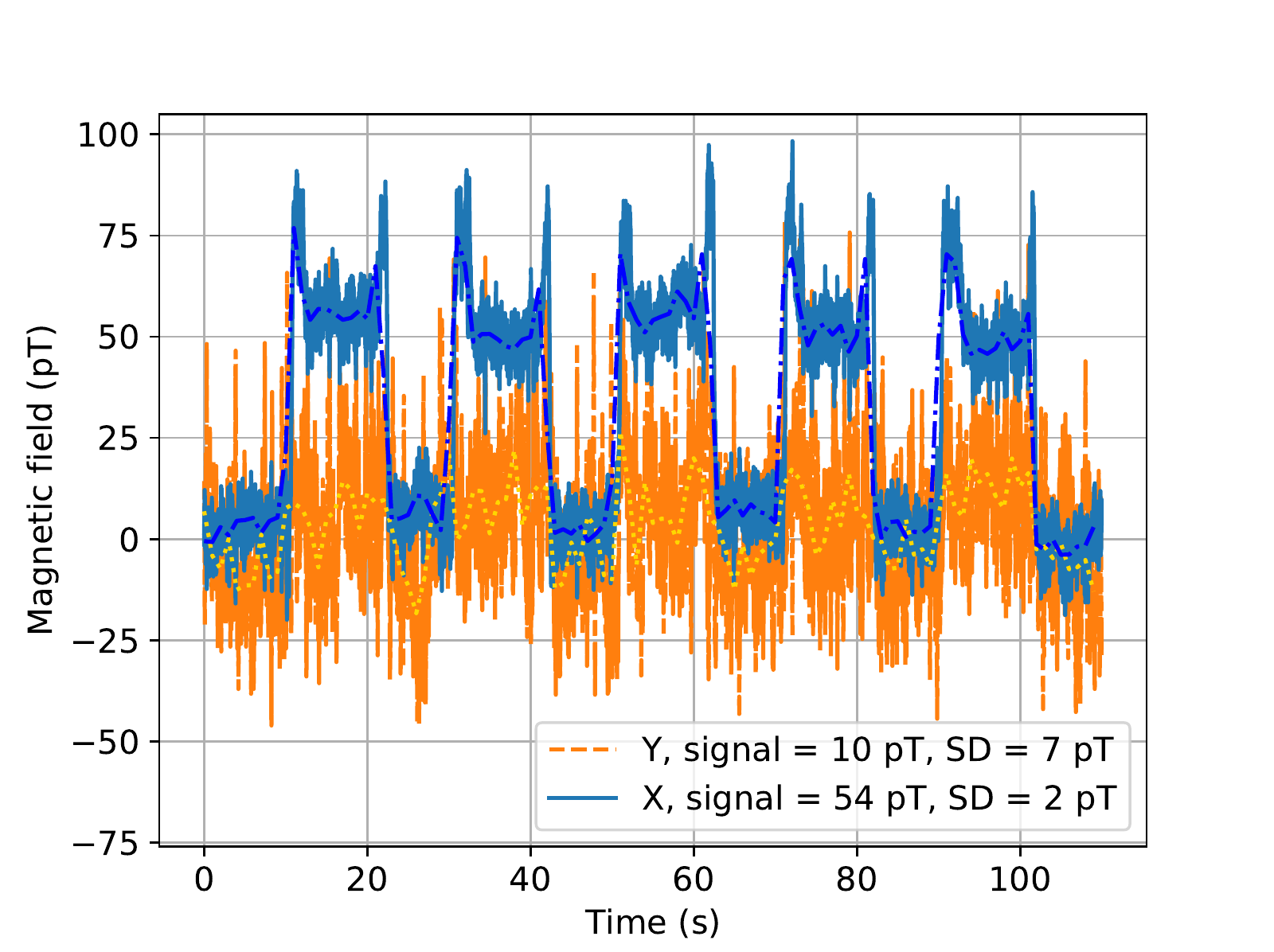}}
\subfigure[]{\label{fig:1point5cm_diameter}\includegraphics[width=8.5cm]{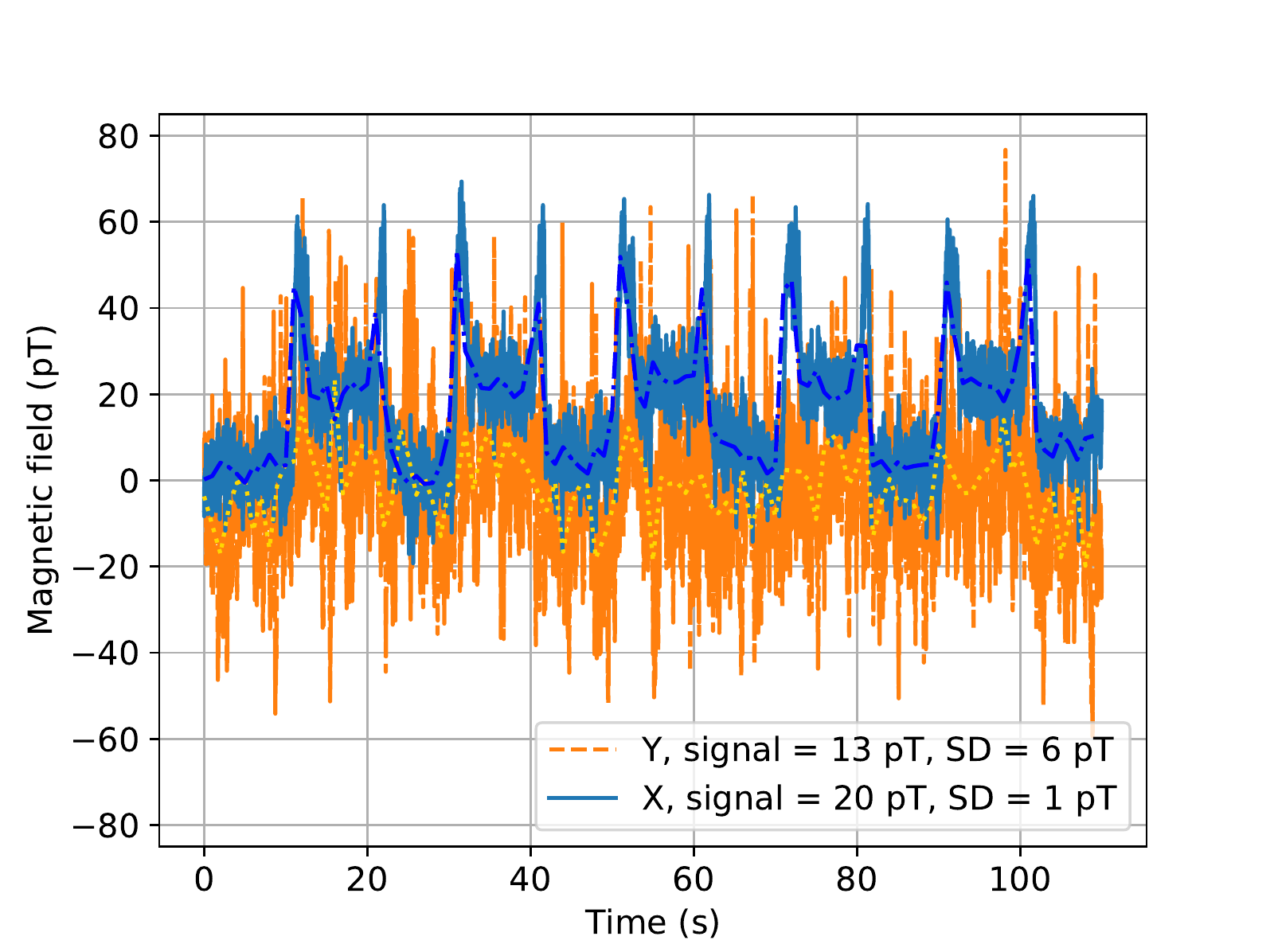}}
\caption{\label{fig:TimeTraces_AllDiameters} Time traces of the eddy current measurements for (a) 5~cm, (b) 3~cm, (c) 2~cm, (d) 1.5~cm diameter Al disks, all with 4~mm thicknesses. The disks were placed 26.4~cm from the excitation coil and 23.9~cm from the vapour cell.}
\end{figure*}

Time traces of the eddy current measurements are shown in Fig.~\ref{fig:TimeTraces_AllDiameters}, when Al disks with 5~cm, 3~cm, 2~cm and 1.5~cm diameters are placed 26.4~cm from the excitation coil and 23.9~cm from the vapour cell. The spikes in the time traces arise when the disk is in the process of being placed in front of the excitation coil. The signal then remains stable, before the object is then removed. The data in the stable region was used for the calculation of the signal size.

\section{Induced magnetic field from a conductive sphere}
\label{sec:Appendix_CalculatingBec_NonMagneticSphere}

We now calculate the expected induced magnetic field for a conductive, non-magnetic solid sphere positioned in between an excitation coil and magnetometer (see inset in Fig.~\ref{fig:PhotoOfSetup}).
The primary magnetic field from the excitation coil (positioned at $x=0$) at the position of the OPM $x=r+r'$ is equal to 
\begin{equation}
    B_{1}(x=r+r') = \frac{\mu_{0}m}{2\pi (r+r')^{3}},
    \label{eq:B1}
\end{equation}
where $m$ is the magnetic moment of the excitation coil, $r$ is the distance from the excitation coil to the sphere with radius $a$ and $r'$ is the distance from the sphere to the OPM. Eq.~\ref{eq:B1} is the on-axis field for a magnetic dipole and is true when $r+r'\gg R_{c}$, where $R_{c}$ is the radius of the excitation coil.

The secondary magnetic field at the position of the OPM $B_{\text{ec}}(x=r+r')$ is calculated to be
\begin{equation}
    B_{\text{ec}}(x=r+r') = \frac{\mu_{0}m_{\text{ec}}}{2\pi r'^{3}},
\end{equation}
where $m_{\text{ec}}$ is the induced magnetic moment in the sphere and $r'$ is the distance from the sphere to the OPM. For a non-magnetic, conductive sphere, $m_{\text{ec}}$ is given by \cite{bidinosti_chapple_hayden_2007, honke_bidinosti_2018}

\begin{equation}
    m_{\text{ec}} = \frac{2\pi a^{3} B_{1}(x=r)}{\mu_{0}}\frac{j_{2}(ka)}{j_{0}(ka)},
    \label{eq:bidinosti_inducedmagneticmoment}
\end{equation}
where
\begin{equation}
    j_{2}(x) = \left(\frac{3}{x^{3}}-\frac{1}{x}\right)\sin x - \frac{3}{x^{2}}\cos x,
\end{equation}
\begin{equation}
    j_{0}(x) = \frac{\sin x}{x},
\end{equation}
and
\begin{equation}
    k = \sqrt{\mu\epsilon\omega^{2} + i \mu \sigma \omega},
\end{equation}
where $k$ is the propagation constant, $\mu=\mu_{0}\mu_{r}$ and $\epsilon=\epsilon_{r}\epsilon_{0}$. The propagation constant can be approximated to be $k\sim \sqrt{i\mu_{0}\sigma\omega}$ for this experiment. In the high-frequency limit where $\delta \ll a$, as is the case throughout this paper, 
\begin{equation}
    \frac{j_{2}(ka)}{j_{0}(ka)} \rightarrow -1
\end{equation}
and hence the secondary magnetic field at the position of the OPM $B_{\text{ec}}(x=r+r')$ is calculated to be
\begin{equation}
    B_{\text{ec}}(x=r+r') = -\frac{a^{3}B_{1}(x=r)}{r'^{3}}.
\end{equation}
The ratio of the induced magnetic field to the primary magnetic field at the position of the OPM is calculated to be
\begin{equation}
    \frac{B_{\text{ec}}(x=r+r')}{B_{1}(x=r+r')} = -\frac{a^{3}}{r'^{3}}\frac{B_{1}(x=r)}{B_{1}(x=r+r')} = -\frac{a^{3}}{r'^{3}}\frac{(r+r')^{3}}{r^{3}},
    \label{eq:figureofmerit}
\end{equation}
which is chosen as the figure of merit in this paper for the remote detection of conductive objects. If the object is close to the excitation coil with radius $R_{c}$, however, 
\begin{equation}
    B_{1}(x=r) = \frac{\mu_{0}m}{2\pi (r^{2}+R_{c}^{2})^{3/2}} \label{eq:b1_close_to_excitation_coil}
\end{equation}
and hence 
\begin{equation}
    \frac{B_{\text{ec}}(x=r+r')}{B_{1}(x=r+r')} = -\frac{a^{3}}{r'^{3}}\frac{(r+r')^{3}}{(r^{2}+R_{c}^{2})^{3/2}}.
    \label{eq:figureofmerit_b1closetoexcitationcoil}
\end{equation}

If the sphere is exactly halfway between the OPM and the excitation coil i.e. $r=r'$ (and $r\gg R_{c}$), then Eq.~\ref{eq:figureofmerit} simplifies to
\begin{equation}
    \frac{B_{\text{ec}}(x=r+r')}{B_{1}(x=r+r')} = -\frac{8a^{3}}{r^{3}}.
\end{equation}

\section{Comparison with co-located excitation coil and magnetometer}
\label{sec:Appendix_AdvantageDifferentialTechnique}
We now consider the situation where the excitation coil and the magnetometer are co-located. The primary field $B_{1}(x=0)$ at the position of the OPM $x=0$ is in this case given by 
\begin{equation}
    B_{1}(x=0) = \frac{\mu_{0}m}{2R_{c}^{3}},
\end{equation}
where $m=\pi R_{c}^{2}nI$, $n$ is the number of windings and $I$ is the current flowing through the coil.

Alternatively, if the OPM is placed on the other side of the object (i.e. a distance $2r$ away from the primary coil assuming the object is centred between the primary coil and OPM), then $B_{1}(x=2r)$ will be given by 
\begin{equation}
    B_{1}(x=2r) = \frac{\mu_{0}m}{2(R_{c}^{2}+(2r)^{2})^{3/2}}.
\end{equation}
In both cases, the induced magnetic field $B_{\text{ec}}$ at the position of the OPM is the same.
As previously discussed , it is important to reduce the effect of $B_{1}$ on the OPM. We can compare the primary magnetic field at the OPM position for the two cases
\begin{equation}
    \frac{B_{1}(x=2r)}{B_{1}(x=0)} = \frac{R_{c}^{3}}{(R_{c}^{2}+4r^{2})^{3/2}}=\frac{R_{c}^{3}}{8r^{3}(R_{c}^{2}/(4r^{2})+1)^{3/2}}.
\end{equation}
In the limit where the object is placed far from the excitation coil ($r \gg R_{c}$), this expression simplifies to
\begin{equation}
    \frac{B_{1}(x=2r)}{B_{1}(x=0)} = \frac{R_{c}^{3}}{8r^{3}}. 
\end{equation}
Inserting the relevant numbers for our setup ($r\sim25$~cm and $R_c\sim6$~cm) we calculate $B_{1}(x=2r)/B_{1}(x=0)\sim0.002$.
By placing the excitation coil and the OPM on opposite sides of the object, the primary magnetic field is orders of magnitude smaller at the OPM position. This configuration will therefore enable much larger detection distances compared to if the excitation coil and the OPM were co-located.

\section*{References}

%\bibliography{references}

\end{document}